\documentclass[12pt,a4paper]{article}
\usepackage{amssymb}

\usepackage{amsmath}

\newcounter{resultnum}[section]
\newtheorem{conclusion}{Conclusion}[section]

\newcounter{conclusionnum}[section]

\newcounter{conditionnum}[section]

\newcounter{conjecturenum}[section]
\newtheorem{example}{Example}[section]

\newcounter{examplenum}[section]

\newcounter{exercisenum}[section]
\newtheorem{lemma}{Lemma}[section]

\newcounter{lemmanum}[section]

\newcounter{notationnum}[section]
\newtheorem{theorem}{Theorem}[section]

\newcounter{theoremnum}[section]
\newtheorem{definition}{Definition}[section]

\newcounter{definitionnum}[section]
\newtheorem{corollary}{Corollary}[section]

\newcounter{corollarynum}[section]
\newtheorem{remark}{Remark}[section]

\newcounter{remarknum}[section]
\newtheorem{proposition}{Proposition}[section]

\newcounter{propositionnum}[section]

\newcounter{acknowledgementnum}[section]

\newcounter{algorithmnum}[section]

\newcounter{axiomnum}[section]

\newcounter{casenum}[section]

\newcounter{claimnum}[section]

\newcounter{summarynum}[section]

\newcounter{problemnum}[section]
\newenvironment{proof}[1][]{\textbf{Proof.} }{}

\begin{document}

\title{Curve Flows and Solitonic Hierarchies Generated by (Semi) Riemannian Metrics }
\date{August 08, 2006}
\author{Sergiu I. Vacaru\thanks{%
Address for correspondence:\ 108-1490 Eglinton Av. West, Toronto, Canada M6E
2G5. \newline
{\ }\quad E--mails: svacaru@brocku.ca, sergiu$_{-}$vacaru@yahoo.com }
{\ } \and \textsl{\ Department of Mathematics, Brock University} \and
\textsl{St. Catharines, Ontario, Canada L2S\ 3A1 } }
\maketitle

\begin{abstract}
We investigate bi--Hamiltonian structures and related mKdV hierarchy of
solitonic equations generated by (semi) Riemannian metrics and curve flow of
non--stretching curves. The corresponding nonholonomic tangent space
geometry is defined by canonically induced nonlinear connections, Sasaki
type metrics and linear connections. One yields couples of generalized
sine--Gordon equations when the corresponding geometric curve flows result
in hierarchies on the tangent bundle described in explicit form by
nonholonomic wave map equations and mKdV analogs of the Schr\"{o}dinger map
equation.


\end{abstract}



\section{ Introduction}

In recent years, the differential geometry of plane and space curves is
receiving considerable attention in the theory of nonlinear partial
differential equations and applications to modern physics \cite%
{chou1,chou2,mbsw,sw,aw}. One proved that curve flows on Riemannian spaces
of constant curvature are described geometrically by hierarchies defined by
wave map equations and mKdV analogs of Schr\"{o}dinger map equation. The
main results on vector generalizations of KdV and mKdV equations and the
geometry of their Hamiltonian structures are summarized in Refs. \cite%
{ath,saw,serg}, see also a recent work in \cite{fours,wang}.

In \cite{anc1,anc2}, the flows of non--stretching curves were analyzed using
moving parallel frames and associated frame connection 1--forms in a
symmetric spaces $M=G/SO(n)$ and the structure equations for torsion and
curvature encoding $O(n-1)$--invariant bi--Hamiltonian operators.\footnote{$%
G $ is a compact semisimple Lie group with an involutive automorphism that
leaves fixed a Lie subgroup $SO(n)\subset G,$ for $n\geq 2$} It was shown
that the bi--Hamiltonian operators produce hierarchies of integrable flows
of curves in which the frame components of the principal normal along the
curve satisfy $O(n-1)$--soliton equations. The crucial condition for such
constructions is the fact that the frame curvature matrix is constant on the
curved manifolds like $M=G/SO(n).$ The approach was developed into a
geometric formalism mapping regular Lagrange mechanical systems into
bi--Hamiltonian structures and related solitonic equations \cite{avw},
following certain methods elaborated in the geometry of generalized Finsler
and Lagrange spaces \cite{ma1,ma2,bej} and nonholonomic manifolds with
applications in modern gravity \cite{bejf,vncg,vsgg}.

The aim of this paper is to prove that solitonic hierarchies can be
generated by any (semi) Riemannian metric $g_{ij}$ on a manifold $V$ of
dimension $\dim V=n\geq 2$ if the the geometrical objects are lifted in the
total space of the tangent bundle $TV,$ or of a vector bundle $\mathcal{E}%
=(M,\pi ,E),$ $\dim E=m\geq n,$ by defining such frame transforms when
constant matrix curvatures are defined canonically with respect to certain
classes of preferred systems of reference.

The paper is organized as follows:

In section 2 we outline the geometry of vector bundles provided with
nonlinear connection. We emphasize the possibility to define fundamental
geometric objects induced by a (semi) Riemannian metric on the base space
when the Riemannian curvature tensor has constant coefficients with respect
to a preferred nonholonomic basis.

In section 3 we consider curve flows on nonholonomic vector bundles. We
sketch an approach to classification of such spaces defined by conventional
horizontal and vertical symmetric (semi) Riemannian subspaces and provided
with nonholonomic distributions defined by the nonlinear connection
structure. It is constructed a class of nonholonomic Klein spaces for which
the bi--Hamiltonian operators are derived for a canonical distinguished
connection, adapted to the nonlinear connection structure, for which the
distinguished curvature coefficients are constant.

Section 4 is devoted to the formalism of distinguished bi--Hamiltonian
operators and vector soliton equations for arbitrary (semi) Riemannian
spaces. We define the basic equations for nonholonomic curve flows. Then we
consider the properties of cosympletic and sympletic operators adapted to
the nonlinear connection structure. Finally, there are constructed solitonic
hierarchies of bi--Hamiltonian anholonomic curve flows

We conclude the results in section 5. The Appendix contains necessary
definitions and formulas from the geometry of nonholonomic manifolds.

\section{Nonholonomic Structures on Manifolds}

In this section, we prove that for any (semi) Riemannian metric $g_{ij}$ on
a manifold $V$ it is possible to define lifts to the tangent bundle $TV$
provided with canonical nonlinear connection (in brief, N--connection),
Sasaki type metric and canonical linear connection structure. The geometric
constructions will be elaborated in general form for vector bundles.

\subsection{N--connections induced by Riemannian metrics}

Let $\mathcal{E}=(E,\pi , F,M)$ be a (smooth) vector bundle of over base
manifold $M,$ when the dimensions are stated respectively; $\dim M=n$ and $%
\dim E=(n+m),$ for $n\geq 2,$ and $m\geq n$ being the dimension of typical
fiber $F.$ It is defined a surjective submersion $\pi :E\rightarrow M.$ In
any point $u\in E,$ the total space $E$ splits into ''horizontal'', $M_{u},$
and ''vertical'', $F_{u},$ subspaces. We denote the local coordinates in the
form $u=(x,y),$ or $u^{\alpha }=\left( x^{i},y^{a}\right) ,$ with horizontal
indices $i,j,k,\ldots =1,2,\ldots ,n$ and vertical indices $a,b,c,\ldots
=n+1,n+2,\ldots ,n+m.$\footnote{%
In a particular case, we have a tangent bundle $E\mathbf{=}TM,$ when $n=m;$
for such bundles both type of indices run the same values but it is
convenient to distinguish the horizontal and vertical ones by using
different groups of Latin indices.} The summation rule on the same "up" and
"low" indices will be applied.

The base manifold $M$ is provided with a (semi) Riemannian metric, a second
rank tensor of constant signature,\footnote{in physical literature,
one uses the term (pseudo) Riemannian/Euclidean space}
 $h\underline{g}=\underline{g}_{ij}(x)dx^{i}\otimes dx^{j}.$
It is possible to introduce a vertical metric structure%
 $v\underline{g}=\underline{g}_{ab}(x)dy^{a}\otimes dy^{b}$
 by completing the matrix $\underline{g}_{ij}(x)$ diagonally with $\pm 1$
till any nondegenerate second rank tensor $\underline{g}_{ab}(x)$ if $m>n.$
This defines a metric structure $\underline{\mathbf{g}}=[h\underline{g}%
,v\underline{g}]$ (we shall also use the notation $\underline{g}_{\alpha
\beta }=[\underline{g}_{ij},\underline{g}_{ab}])$ on $\mathcal{E}.$ We can
deform the metric structure, $\underline{g}_{\alpha \beta }\rightarrow
g_{\alpha \beta }=[g_{ij},g_{ab}],$ by considering a frame (vielbein)
transform,%
\begin{equation}
g_{\alpha \beta }(x,y)=e_{\alpha }^{~\underline{\alpha }}(x,y)~e_{\beta }^{~%
\underline{\beta }}(x,y)g_{\underline{\alpha }\underline{\beta }}(x),
\label{auxm}
\end{equation}%
where the coefficients $\underline{g}_{\alpha \beta }(x)$ have been written
as $g_{\underline{\alpha }\underline{\beta }}(x).$ The coefficients $%
e_{\alpha }^{~\underline{\alpha }}(x,y)$ will be defined below (see formula (%
\ref{aux4})) from the condition of generating curvature tensors with
constant coefficients with respect to certain preferred systems of reference.

For any $g_{ab}$ from the set $g_{\alpha \beta},$ we can construct an
effective generation function
\begin{equation*}
\mathcal{L}(x,y)=g_{ab}(x,y)y^{a}y^{b}
\end{equation*}%
inducing a vertical metric
\begin{equation}
\tilde{g}_{ab}=\frac{1}{2}\frac{\partial ^{2}\mathcal{L}}{\partial
y^{a}\partial y^{b}}  \label{ehes}
\end{equation}%
which is ''weakly'' regular if $\det |\tilde{g}_{ab}|\neq 0.$ \footnote{%
Similar values, for $e_{\alpha }^{~\underline{\alpha }}=\delta _{\alpha }^{~%
\underline{\alpha }},$ where $\delta _{\alpha }^{~\underline{\alpha }}$ is
the Kronecker symbol, were introduced for the so--called generalized
Lagrange spaces when $\mathcal{L}$ was called the ''absolute energy'' \cite%
{ma1}.}

By straightforward calculations we can prove this result \footnote{%
see Refs. \cite{ma1,ma2} for details of a similar proof; here we note that
in our case, in general, \ $e_{\alpha }^{~\underline{\alpha }}\neq \delta
_{\alpha }^{~\underline{\alpha }}$}:

\begin{theorem}
\label{teleq}The Euler--Lagrange equations on $TM,$
\begin{equation*}
\frac{d}{d\tau }\left( \frac{\partial L}{\partial y^{i}}\right) -\frac{%
\partial L}{\partial x^{i}}=0,
\end{equation*}%
for the \ Lagrangian $L=\sqrt{|\mathcal{L}|},$ where $y^{i}=\frac{dx^{i}}{%
d\tau }$ for a path $x^{i}(\tau )$ on $M,$ depending on parameter $\tau ,$
are equivalent to the ``nonlinear'' geodesic equations
\begin{equation*}
\frac{d^{2}x^{i}}{d\tau ^{2}}+2\widetilde{G}^{i}(x^{k},\frac{dx^{j}}{d\tau }%
)=0
\end{equation*}%
defining paths of a canonical semispray
 $S=y^{i}\frac{\partial }{\partial x^{i}}-2\widetilde{G}^{i}(x,y)\frac{%
\partial }{\partial y^{i}},$
where
\begin{equation*}
2\widetilde{G}^{i}(x,y)=\frac{1}{2}\ \tilde{g}^{ij}\left( \frac{\partial
^{2}L}{\partial y^{i}\partial x^{k}}y^{k}-\frac{\partial L}{\partial x^{i}}%
\right)
\end{equation*}%
with $\tilde{g}^{ij}$ being inverse to (\ref{ehes}).
\end{theorem}

On holds

\begin{conclusion}
For any (semi) Riemannian metric $\underline{g}_{ij}(x)$ on $M,$ we can
associate canonically an effective regular Lagrange mechanics on $TM$ with
the Euler--Lagrange equations transformed into nonlinear (semispray)
geodesic equations.
\end{conclusion}

We denote by $\pi ^{\top }:TE\rightarrow TM$ the differential of map $\pi
:E\rightarrow M$ defined by fiber preserving morphisms of the tangent
bundles $TE$ and $TM.$ The kernel of $\pi ^{\top }$ is just the vertical
subspace $vE$ with a related inclusion mapping $i:vE\rightarrow TE.$

\begin{definition}
A nonlinear connection (N--connection) $\mathbf{N}$ on a vector bundle $%
\mathcal{E}$ \ is defined by the splitting on the left of an exact sequence
\begin{equation*}
0\rightarrow vE\overset{i}{\rightarrow }TE\rightarrow TE/vE\rightarrow 0,
\end{equation*}%
i. e. by a morphism of submanifolds $\mathbf{N:\ \ }TE\rightarrow vE$ such
that $\mathbf{N\circ }i$ is the unity in $vE.$
\end{definition}

In an equivalent form, we can say that a N--connection is defined by a
Whitney sum of conventional horizontal (h) subspace, $\left( hE\right) ,$
and vertical (v) subspace, $\left( vE\right) ,$
\begin{equation}
TE=hE\oplus vE.  \label{whitney}
\end{equation}%
This sum defines a nonholonomic (equivalently, anholonomic, or nonitegrable)
distribution of horizontal and vertical subspaces on $TE\mathbf{.}$ Locally,
a N--connection is defined by its coefficients $N_{i}^{a}(u),$%
\begin{equation*}
\mathbf{N}=N_{i}^{a}(u)dx^{i}\otimes \frac{\partial }{\partial y^{a}}.
\end{equation*}%
The well known class of linear connections consists on a particular subclass
with the coefficients being linear on $y^{a},$ i.e., $N_{i}^{a}(u)=\Gamma
_{bj}^{a}(x)y^{b}.$

\begin{remark}
A bundle space, or a a manifold, is called nonholonomic if it provided with
a nonholonomic distribution (see historical details and summary of results
in \cite{bejf}). In particular case, when the nonholonomic distribution is
of type (\ref{whitney}), such spaces are called N--anholonomic \cite{vsgg}.
\end{remark}

Any N--connection $\mathbf{N}=\left\{ N_{i}^{a}(u)\right\} $ may be
characterized by a N--adapted frame (vielbein) structure $\mathbf{e}_{\nu
}=(e_{i},e_{a}),$ where
\begin{equation}
\mathbf{e}_{i}=\frac{\partial }{\partial x^{i}}-N_{i}^{a}(u)\frac{\partial }{%
\partial y^{a}}\mbox{ and
}e_{a}=\frac{\partial }{\partial y^{a}},  \label{dder}
\end{equation}%
and the dual frame (coframe) structure $\mathbf{e}^{\mu }=(e^{i},\mathbf{e}%
^{a}),$ where
\begin{equation}
e^{i}=dx^{i}\mbox{ and }\mathbf{e}^{a}=dy^{a}+N_{i}^{a}(u)dx^{i}.
\label{ddif}
\end{equation}
In order to preserve a relation with the previous denotations, we note that $%
\mathbf{e}_{\nu }=(\mathbf{e}_{i},e_{a})$ and $\mathbf{e}^{\mu }=(e^{i},%
\mathbf{e}^{a})$ are, respectively, the former ''N--elongated'' partial
derivatives $\delta _{\nu }=\delta /\partial u^{\nu }=(\delta _{i},\partial
_{a})$ and N--elongated differentials $\delta ^{\mu }=\delta u^{\mu
}=(d^{i},\delta ^{a})$ which emphasize that operators (\ref{dder}) and (\ref%
{ddif}) define, correspondingly, certain ``N--elongated'' partial
derivatives and differentials which are more convenient for tensor and
integral calculations on such nonholonomic manifolds.\footnote{%
We shall use ''boldface'' symbols if it would be necessary to emphasize that
any space and/or geometrical objects are provided/adapted to a\
N--connection structure, or with the coefficients computed with respect to
N--adapted frames.}

For any N--connection, we can introduce its N--connection curvature
\begin{equation*}
\mathbf{\Omega }=\frac{1}{2}\Omega _{ij}^{a}\ d^{i}\wedge d^{j}\otimes
\partial _{a},
\end{equation*}%
with the coefficients defined as the Neijenheuse tensor,%
\begin{equation}
\Omega _{ij}^{a}=\mathbf{e}_{[j}N_{i]}^{a}=\mathbf{e}_{j}N_{i}^{a}-\mathbf{e}%
_{i}N_{j}^{a}=\frac{\partial N_{i}^{a}}{\partial x^{j}}-\frac{\partial
N_{j}^{a}}{\partial x^{i}}+N_{i}^{b}\frac{\partial N_{j}^{a}}{\partial y^{b}}%
-N_{j}^{b}\frac{\partial N_{i}^{a}}{\partial y^{b}}.  \label{ncurv}
\end{equation}

The vielbeins (\ref{ddif}) satisfy the nonholonomy (equivalently,
anholonomy) relations
\begin{equation}
\lbrack \mathbf{e}_{\alpha },\mathbf{e}_{\beta }]=\mathbf{e}_{\alpha }%
\mathbf{e}_{\beta }-\mathbf{e}_{\beta }\mathbf{e}_{\alpha }=W_{\alpha \beta
}^{\gamma }\mathbf{e}_{\gamma }  \label{anhrel}
\end{equation}%
with (antisymmetric) nontrivial anholonomy coefficients $W_{ia}^{b}=\partial
_{a}N_{i}^{b}$ and $W_{ji}^{a}=\Omega _{ij}^{a}.$

The geometric objects can be defined in a form adapted to a N--connection
structure, following decompositions being invariant under parallel
transports preserving the splitting (\ref{whitney}). In this case we call
them to be distinguished (by the connection structure), i.e. d--objects. For
instance, a vector field $\mathbf{X}\in T\mathbf{V}$ \ is expressed
\begin{equation*}
\mathbf{X}=(hX,\ vX),\mbox{ \ or \ }\mathbf{X}=X^{\alpha }\mathbf{e}_{\alpha
}=X^{i}\mathbf{e}_{i}+X^{a}e_{a},
\end{equation*}%
where $hX=X^{i}\mathbf{e}_{i}$ and $vX=X^{a}e_{a}$ state, respectively, the
adapted to the N--connection structure horizontal (h) and vertical (v)
components of the vector (which following Refs. \cite{ma1,ma2} is called a
distinguished vector, in brief, d--vector). In a similar fashion, the
geometric objects on $\mathbf{V},$ for instance, tensors, spinors,
connections, ... are called respectively d--tensors, d--spinors,
d--connections if they are adapted to the N--connection splitting (\ref%
{whitney}).

\begin{theorem}
Any (semi) Riemannian metric $\underline{g}_{ij}(x)$ on $M$ induces a
canonical N--connection structure on $TM.$
\end{theorem}

\begin{proof}
We sketch a proof by defining the coefficients of N--connection
\begin{equation}
\tilde{N}_{\ j}^{i}(x,y)=\frac{\partial \tilde{G}^{i}}{\partial y^{j}}
\label{cnlce}
\end{equation}%
where
\begin{eqnarray}
\tilde{G}^{i} &=&\frac{1}{4}\tilde{g}^{ij}\left( \frac{\partial ^{2}\mathcal{%
L}}{\partial y^{i}\partial x^{k}}y^{k}-\frac{\partial \mathcal{L}}{\partial
x^{j}}\right) =\frac{1}{4}\tilde{g}^{ij}g_{jk}\gamma _{lm}^{k}y^{l}y^{m},
\label{aux2} \\
\gamma _{\ lm}^{i} &=&\frac{1}{2}g^{ih}(\partial _{m}g_{lh}+\partial
_{l}g_{mh}-\partial _{h}g_{lm}),\ \partial _{h}=\partial /\partial x^{h},
\notag
\end{eqnarray}%
with $g_{ah}$ and $\tilde{g}_{ij}$ defined respectively by formulas (\ref%
{auxm}) and (\ref{ehes}). $\square$
\end{proof}

The N--adapted operators (\ref{dder}) and (\ref{ddif}) defined by the
N--connection coefficients (\ref{cnlce}) are denoted respectively $\mathbf{%
\tilde{e}}_{\nu }=(\mathbf{\tilde{e}}_{i},e_{a})$ and $\mathbf{\tilde{e}}%
^{\mu }=(e^{i},\mathbf{\tilde{e}}^{a}).$

\subsection{Canonical linear connection and metric structures}

The constructions will be performed on a vector bundle $\mathbf{E}$ provided
with N--connection structure. We shall emphasize the special properties of a
tangent bundle $(TM,\!\pi ,\!M)$ when the linear connection and metric are
induced by a (semi) Riemannian metric on $M.$

\begin{definition}
A distinguished connection (in brief, d--connection) $\mathbf{D}=(h\mathbf{D}%
,v\mathbf{D})$ is a linear connection preserving under parallel transports
the nonholonomic decomposition (\ref{whitney}).
\end{definition}

The N--adapted components $\mathbf{\Gamma }_{\ \beta \gamma }^{\alpha }$ of
a d--connection $\mathbf{D}_{\alpha }=(\mathbf{e}_{\alpha }\rfloor \mathbf{D}%
)$ are defined by the equations
\begin{equation}
\mathbf{D}_{\alpha }\mathbf{e}_{\beta }=\mathbf{\Gamma }_{\ \alpha \beta
}^{\gamma }\mathbf{e}_{\gamma },\mbox{\ or \ }\mathbf{\Gamma }_{\ \alpha
\beta }^{\gamma }\left( u\right) =\left( \mathbf{D}_{\alpha }\mathbf{e}%
_{\beta }\right) \rfloor \mathbf{e}^{\gamma }.  \label{dcon1}
\end{equation}%
The N--adapted splitting into h-- and v--covariant derivatives is stated by
\begin{equation*}
h\mathbf{D}=\{\mathbf{D}_{k}=\left( L_{jk}^{i},L_{bk\;}^{a}\right) \},%
\mbox{
and }\ v\mathbf{D}=\{\mathbf{D}_{c}=\left( C_{jk}^{i},C_{bc}^{a}\right) \},
\end{equation*}%
where, by definition,
$L_{jk}^{i}=\left( \mathbf{D}_{k}\mathbf{e}_{j}\right) \rfloor e^{i},$
$L_{bk}^{a}=\left( \mathbf{D}_{k}e_{b}\right) \rfloor \mathbf{e}^{a},$
$C_{jc}^{i}=\left( \mathbf{D}_{c}\mathbf{e}_{j}\right) \rfloor e^{i},$
$C_{bc}^{a}=\left( \mathbf{D}_{c}e_{b}\right) \rfloor \mathbf{e}^{a}.$
 The components $\mathbf{\Gamma }_{\ \alpha \beta }^{\gamma }=\left(
L_{jk}^{i},L_{bk}^{a},C_{jc}^{i},C_{bc}^{a}\right) $ completely define a
d--connection $\mathbf{D}$ on $\mathbf{E}.$

The simplest way to perform N--adapted computations is to use differential
forms. For instance, starting with the d--connection 1--form,
\begin{equation}
\mathbf{\Gamma }_{\ \beta }^{\alpha }=\mathbf{\Gamma }_{\ \beta \gamma
}^{\alpha }\mathbf{e}^{\gamma },  \label{dconf}
\end{equation}%
with the coefficients defined with respect to N--elongated frames (\ref{ddif}%
) and (\ref{dder}), the torsion of a d--connection,
\begin{equation}
\mathcal{T}^{\alpha }\doteqdot \mathbf{De}^{\alpha }=d\mathbf{e}^{\alpha
}+\Gamma _{\ \beta }^{\alpha }\wedge \mathbf{e}^{\beta },  \label{tors}
\end{equation}
is characterized by (N--adapted) d--torsion components,
\begin{eqnarray}
T_{\ jk}^{i} &=&L_{\ jk}^{i}-L_{\ kj}^{i},\ T_{\ ja}^{i}=-T_{\ aj}^{i}=C_{\
ja}^{i},\ T_{\ ji}^{a}=\Omega _{\ ji}^{a},\   \notag \\
T_{\ bi}^{a} &=&-T_{\ ib}^{a}=\frac{\partial N_{i}^{a}}{\partial y^{b}}-L_{\
bi}^{a},\ T_{\ bc}^{a}=C_{\ bc}^{a}-C_{\ cb}^{a}.  \label{dtors}
\end{eqnarray}%
For d--connection structures on $TM,$ we have to identify indices in the
form $i\leftrightarrows a,j\leftrightarrows b,...$ and the components of N--
and d--connections, for instance, $N_{i}^{a}\leftrightarrows N_{i}^{j}$ and $%
L_{\ jk}^{i}\leftrightarrows L_{\ bk}^{a},C_{\ ja}^{i}\leftrightarrows C_{\
ca}^{b}\leftrightarrows C_{\ jk}^{i}.$

\begin{definition}
A distinguished metric (in brief, d--metric) on a vector bundle $\mathbf{E}$
is a usual second rank metric tensor $\mathbf{g=}g\mathbf{\oplus _{N}}h$
equivalently
\begin{equation}
\mathbf{g}=\ g_{ij}(x,y)\ e^{i}\otimes e^{j}+\ h_{ab}(x,y)\ \mathbf{e}%
^{a}\otimes \mathbf{e}^{b},  \label{m1}
\end{equation}%
adapted to the N--connection decomposition (\ref{whitney}).
\end{definition}

From the class of arbitrary d--connections $\mathbf{D}$ on $\mathbf{V,}$ one
distinguishes those which are metric compatible (metrical) satisfying the
condition%
\begin{equation}
\mathbf{Dg=0}  \label{metcomp}
\end{equation}%
including all h- and v-projections
$D_{j}g_{kl}=0,$ $D_{a}g_{kl}=0,$ $D_{j}h_{ab}=0,$ $D_{a}h_{bc}=0.$
For d--metric structures on $\mathbf{V\simeq }TM,$ with $g_{ij}=h_{ab},$ the
condition of vanishing ''nonmetricity'' (\ref{metcomp}) transform into%
\begin{equation}
h\mathbf{D(}g\mathbf{)=}0\mbox{\  and\ }v\mathbf{D(}h\mathbf{)=}0,
\label{metcompt}
\end{equation}%
i.e. $D_{j}g_{kl}=0$ and $D_{a}g_{kl}=0.$

For any metric structure $\mathbf{g}$ on a manifold, there is the unique
metric compatible and torsionless Levi Civita connection $\nabla $ for which
$\ ^{\nabla }\mathcal{T}^{\alpha }=0$ and $\nabla \mathbf{g=0.}$ This
connection is not a d--connection because it does not preserve under
parallelism the N--connection splitting (\ref{whitney}). One has to consider
less constrained cases, admitting nonzero torsion coefficients, when a
d--connection is constructed canonically for a d--metric structure. A simple
minimal metric compatible extension of $\nabla $ is that of canonical
d--connection $\widehat{\mathbf{D}}$ which is metric compatible, with $T_{\
jk}^{i}=0$ and $T_{\ bc}^{a}=0$ but $T_{\ ja}^{i},T_{\ ji}^{a}$ and $T_{\
bi}^{a}$ are not zero, see (\ref{dtors}). The coefficient formulas for such
connections are given in Appendix, see (\ref{candcon}) and related
discussion.

\begin{lemma}
Any (semi) Riemannian metric $\underline{g}_{ij}(x)$ on a manifold $M$
induces a canonical d--metric structure on $TM,$
\begin{equation}
\mathbf{\tilde{g}}=\tilde{g}_{ij}(x,y)\ e^{i}\otimes e^{j}+\ \tilde{g}%
_{ij}(x,y)\ \mathbf{\tilde{e}}^{i}\otimes \mathbf{\tilde{e}}^{j},
\label{slme}
\end{equation}%
where $\mathbf{\tilde{e}}^{i}$ are elongated as in (\ref{ddif}), but with $%
\tilde{N}_{\ j}^{i}$ from (\ref{cnlce}).
\end{lemma}

\begin{proof}
This construction is similar to that of lifting of the so--called Sasaki
metric \cite{yano}, but using the coefficients $\tilde{g}_{ij}$ (\ref{ehes}).%
$\square $
\end{proof}

\begin{proposition}
There are canonical d--connections on $TM$ induced by a (semi) Riemannian
metric $\underline{g}_{ij}(x)$ on $M.$
\end{proposition}

\begin{proof}
We can construct an example in explicit form by introducing $\tilde{g}_{ij}$
and $\tilde{g}_{ab}$ in formulas (\ref{candcontm}), see Appendix, in order
to compute the coefficients $\tilde{\Gamma}_{\ \beta \gamma }^{\alpha }=(%
\tilde{L}_{\ jk}^{i},\tilde{C}_{bc}^{a}).\square $
\end{proof}

From the above Lemma and Proposition, one follows the proof of

\begin{theorem}
Any (semi) Riemannian metric $\underline{g}_{ij}(x)$ on $M$ induces a
nonholonomic (semi) Riemannian structure on $TM.$
\end{theorem}

We note that the induced Riemannian structure is nonholonomic because there
is also a nonholonomic distribution (\ref{whitney}) defining $\tilde{N}_{\
j}^{i}.$ The corresponding curvature curvature tensor $\widetilde{R}_{\
\beta \gamma \tau }^{\alpha }=\{\widetilde{R}_{\ hjk}^{i},\tilde{P}_{\
jka}^{i},\tilde{S}_{\ bcd}^{a}\}$ can be computed by introducing
respectively the values $\tilde{g}_{ij},\tilde{N}_{\ j}^{i}$ and $\widetilde{%
\mathbf{e}}_{k}$ into formulas (\ref{dcurvtb}) from Appendix, for defined $%
\tilde{\Gamma}_{\ \beta \gamma }^{\alpha }=(\tilde{L}_{\ jk}^{i},\tilde{C}%
_{bc}^{a}).$ Here one should be noted that the constructions on $TM$ depend
on arbitrary vielbein coefficients $e_{\alpha }^{~\underline{\alpha }}(x,y)$
in (\ref{auxm}). We can restrict such sets of coefficients in order to
generate various particular classes of (semi) Riemannian geometries on $TM,$
for instance, in order to generate symmetric Riemannian spaces with constant
curvature, see Refs. \cite{helag,kob,sharpe}.

\begin{corollary}
There are lifts of a (semi) Riemannian metric $\underline{g}_{ij}(x)$ on $M,$
$\dim M=n,$ generating a Riemannian structure on $TM$ with the curvature
coefficients of the canonical d--connection coinciding (with respect to
N--adapted bases) to those for a Riemannian space of constant curvature of
dimension $n+n.$
\end{corollary}

\begin{proof}
For a given set $\underline{g}_{ij}(x)$ on $M,$ we chose such coefficients $%
e_{\alpha }^{~\underline{\alpha }}(x,y)=\left\{ e_{a}^{~\underline{a}%
}(x,y)\right\} $ in (\ref{auxm}) that
\begin{equation*}
g_{ab}(x,y)=e_{a}^{~\underline{a}}(x,y)~e_{b}^{~\underline{b}}(x,y)g_{%
\underline{a}\underline{b}}(x)
\end{equation*}%
results in (\ref{ehes}) of type%
\begin{equation}
\tilde{g}_{ef}=\frac{1}{2}\frac{\partial ^{2}\mathcal{L}}{\partial
y^{e}\partial y^{f}}=\frac{1}{2}\frac{\partial ^{2}(e_{a}^{~\underline{a}%
}~e_{b}^{~\underline{b}}y^{a}y^{b})}{\partial y^{e}\partial y^{f}}g_{%
\underline{a}\underline{b}}(x)=\ \mathring{g}_{ef},  \label{aux4}
\end{equation}%
where $\ \mathring{g}_{ab}$ is the metric of a symmetric Riemannian space
(of constant curvature). Considering a prescribed $~\ \mathring{g}_{ab},$ we
have to integrate two times on $y^{e}$ in order to find any solution for $%
e_{a}^{~\underline{a}}$ defining a frame structure in the vertical subspace.
The next step is to construct the d--metric $~\ \mathring{g}_{\alpha \beta }=[\
\mathring{g}_{ij},\ \mathring{g}_{ab}]$ of type (\ref{slme}), in our case,
with respect to a nonholonomic base elongated by $~\widetilde{\mathring{N}}%
_{\ j}^{i},$ generated by $\underline{g}_{ij}(x)$ and $\tilde{g}_{ef}=%
\mathring{g}_{ab},$ like in (\ref{cnlce}) and (\ref{aux2}). This defines a
constant curvature Riemannian space of dimension $n+n.$ The coefficients of
the canonical d--connection, which in this case coincide with those for the
Levi Civita connection, and the coefficients of the Riemannian curvature can
be computed respectively by introducing $\tilde{g}_{ef}=\ \mathring{g}_{ab}$
in formulas (\ref{candcontm}) and (\ref{dcurvtb}), see Appendix. Finally, we
note that the induced symmetric Riemannian space contains additional
geometric structures like the N--connection and anholonomy coefficients $%
W_{\alpha \beta }^{\gamma },$ see (\ref{anhrel}).$\square $
\end{proof}

There are various possibilities to generate on $TM$ nonholonomic Riemannian
structures from a given set $\underline{g}_{ij}(x)$ on $M.$ They result in
different geometrical and physical models. In this work, we emphasize the
possibility of generating spaces with constant curvature because for such
symmetric spaces it was elaborated a bi--Hamiltonian approach to solitonic
hierarchies.

\begin{example}
The simplest example when a Riemannian structure with constant matrix
curvature coefficients is generated on $TM$ is to consider a d--metric
induced by $\tilde{g}_{ij}=\delta _{ij},$ i.e.%
\begin{equation}
\mathbf{\tilde{g}}_{[E]}=\delta _{ij}e^{i}\otimes e^{j}+\ \delta _{ij}\
\mathbf{\tilde{e}}^{i}\otimes \mathbf{\tilde{e}}^{j},  \label{clgs}
\end{equation}%
with $\mathbf{\tilde{e}}^{i}$ defined by $\tilde{N}_{\ j}^{i}$ in their turn
defined by a given set $\underline{g}_{ij}(x)$ on $M.$
\end{example}

It should be noted that the metric (\ref{clgs}) is generic off--diagonal
with respect to a coordinate bases because, in general, the anholonomy
coefficients from (\ref{anhrel}) are not zero. This way, we model on $TM$ a
nonholonomic Euclidean space with vanishing curvature coefficients of the
canonical d--connection (it can be verified by introducing respectively the
constant coefficients of metric (\ref{clgs}) into formulas (\ref{candcontm})
and (\ref{dcurvtb})). We note that the conditions of Theorem \ref{teleq} are
not satisfied by the d--metric (\ref{clgs}) (the coefficients $\tilde{g}%
_{ij}=\delta _{ij}$ are not defined as in (\ref{ehes})), so we can not
relate directly a geometrical mechanics model for such constructions.

There is an important generalization:

\begin{example}
We can consider $\mathcal{L}$ as a hypersurface in $TM$ for which the matrix
$\partial ^{2}\mathcal{L}/\partial y^{a}\partial y^{b}$ (i.e. the Hessian,
following the analogy with Lagrange mechanics and field theory) is constant
and nondegenerate. This states that $\tilde{g}_{ij}=const$ which results in
zero curvature coefficients for the canonical d--connection induced by $%
\underline{g}_{ij}(x)$ on $M.$
\end{example}

Finally, in this section, we note that a number of geometric ideas and
methods applied in this section were considered in the approaches to the
geometry of nonholonomic spaces and generalized Finsler--Lagrange geometry
elaborated by the schools of G. Vranceanu and R. Miron and by A. Bejancu in
Romania \cite{vr1,vr2,ma1,ma2,bej,bejf}. We emphasize that this way it is
possible to construct geometric models with metric compatible linear
connections which is important for elaborating standard approaches
compatible with modern (non)commutative gravity and string theory \cite%
{vncg,vsgg}. For Finsler spaces with nontrivial metricity, for instance, for
those defined by the the Berwald and Chern connections, see details in \cite%
{bcs}, the physical theories with local anisotropy are not imbedded in the
class of standard models.

\section{Curve Flows and Anholonomic Constraints}

We formulate the geometry of curve flows adapted to the nonlinear connection
structure.

\subsection{Non--stretching and N--adapted curve flows}

Let us consider a vector bundle $\mathcal{E}=(E,\pi ,F,M),$ $\dim E=$ $n+m$
(in a particular case, $E=TM,$ when $m=n)$ provided with d--metric $\mathbf{g%
}=[g,h]$ (\ref{m1}) and N--connection $N_{i}^{a}$ (\ref{whitney})
structures. A non--stretching curve $\gamma (\tau ,\mathbf{l})$ on $\mathbf{%
V,}$ where $\tau $ is a parameter and $\mathbf{l}$ is the arclength of the
curve on $\mathbf{V,}$ is defined with such evolution d--vector $\mathbf{Y}%
=\gamma _{\tau }$ and tangent d--vector $\mathbf{X}=\gamma _{\mathbf{l}}$
that $\mathbf{g(X,X)=}1\mathbf{.}$ A such curve $\gamma (\tau ,\mathbf{l})$
swept out a two--dimensional surface in $T_{\gamma (\tau ,\mathbf{l})}%
\mathbf{V}\subset T\mathbf{V.}$

We shall work with N--adapted bases (\ref{dder}) and (\ref{ddif}) and the
connection 1--form $\mathbf{\Gamma }_{\ \beta }^{\alpha }=\mathbf{\Gamma }%
_{\ \beta \gamma }^{\alpha }\mathbf{e}^{\gamma }$ with the coefficients $%
\mathbf{\Gamma }_{\ \beta \gamma }^{\alpha }$ for the canonical
d--connection operator $\mathbf{D}$ (\ref{candcon}) (see Appendix) acting in
the form%
\begin{equation}
\mathbf{D}_{\mathbf{X}}\mathbf{e}_{\alpha }=(\mathbf{X\rfloor \Gamma }%
_{\alpha \ }^{\ \gamma })\mathbf{e}_{\gamma }\mbox{ and }\mathbf{D}_{\mathbf{%
Y}}\mathbf{e}_{\alpha }=(\mathbf{Y\rfloor \Gamma }_{\alpha \ }^{\ \gamma })%
\mathbf{e}_{\gamma },  \label{part01}
\end{equation}%
where ''$\mathbf{\rfloor "}$ denotes the interior product and the indices
are lowered and raised respectively by the d--metric $\mathbf{g}_{\alpha
\beta }=[g_{ij},h_{ab}]$ and its inverse $\mathbf{g}^{\alpha \beta
}=[g^{ij},h^{ab}].$ We note that $\mathbf{D}_{\mathbf{X}}=\mathbf{X}^{\alpha
}\mathbf{D}_{\alpha }$ is the covariant derivation operator along curve $%
\gamma (\tau ,\mathbf{l}).$ It is convenient to fix the N--adapted frame to
be parallel to curve $\gamma (\mathbf{l})$ adapted in the form
\begin{eqnarray}
e^{1} &\doteqdot &h\mathbf{X,}\mbox{ for }i=1,\mbox{ and }e^{\widehat{i}},%
\mbox{ where }h\mathbf{g(}h\mathbf{X,}e^{\widehat{i}}\mathbf{)=}0,
\label{curvframe} \\
\mathbf{e}^{n+1} &\doteqdot &v\mathbf{X,}\mbox{ for }a=n+1,\mbox{ and }%
\mathbf{e}^{\widehat{a}},\mbox{ where }v\mathbf{g(}v\mathbf{X,\mathbf{e}}^{%
\widehat{a}}\mathbf{)=}0,  \notag
\end{eqnarray}%
for $\widehat{i}=2,3,...n$ and $\widehat{a}=n+2,n+3,...,n+m.$ For such
frames, the covariant derivative of each ''normal'' d--vectors $\mathbf{e}^{%
\widehat{\alpha }}$ results into the d--vectors adapted to $\gamma (\tau ,%
\mathbf{l}),$
\begin{eqnarray}
\mathbf{D}_{\mathbf{X}}e^{\widehat{i}} &\mathbf{=}&\mathbf{-}\rho ^{\widehat{%
i}}\mathbf{(}u\mathbf{)\ X}\mbox{ and }\mathbf{D}_{h\mathbf{X}}h\mathbf{X}%
=\rho ^{\widehat{i}}\mathbf{(}u\mathbf{)\ \mathbf{e}}_{\widehat{i}},
\label{part02} \\
\mathbf{D}_{\mathbf{X}}\mathbf{\mathbf{e}}^{\widehat{a}} &\mathbf{=}&\mathbf{%
-}\rho ^{\widehat{a}}\mathbf{(}u\mathbf{)\ X}\mbox{ and }\mathbf{D}_{v%
\mathbf{X}}v\mathbf{X}=\rho ^{\widehat{a}}\mathbf{(}u\mathbf{)\ }e_{\widehat{%
a}},  \notag
\end{eqnarray}%
which holds for certain classes of functions $\rho ^{\widehat{i}}\mathbf{(}u%
\mathbf{)}$ and $\rho ^{\widehat{a}}\mathbf{(}u\mathbf{).}$ The formulas (%
\ref{part01}) and (\ref{part02}) are distinguished into h-- and
v--components for $\mathbf{X=}h\mathbf{X}+v\mathbf{X}$ and $\mathbf{D=(}h%
\mathbf{D},v\mathbf{D)}$ for $\mathbf{D=\{\Gamma }_{\ \alpha \beta }^{\gamma
}\},h\mathbf{D}=\{L_{jk}^{i},L_{bk}^{a}\}$ and $v\mathbf{D=\{}%
C_{jc}^{i},C_{bc}^{a}\}.$

Along $\gamma (\mathbf{l}),$ we can move differential forms in a parallel
N--adapted form. For instance, $\mathbf{\Gamma }_{\ \mathbf{X}}^{\alpha
\beta }\doteqdot \mathbf{X\rfloor \Gamma }_{\ }^{\alpha \beta }.$ The
algebraic characterization of such spaces, can be obtained if we perform a
frame transform preserving the decomposition (\ref{whitney}) to an
orthonormalized basis $\mathbf{e}_{\alpha ^{\prime }},$ when
\begin{equation}
\mathbf{e}_{\alpha }\rightarrow A_{\alpha }^{\ \alpha ^{\prime }}(u)\
\mathbf{e}_{\alpha ^{\prime }},  \label{orthbas}
\end{equation}%
called orthonormal d--basis. In this case, the coefficients of the d--metric
(\ref{m1}) transform into the Euclidean ones\textbf{\ }$\mathbf{g}_{\alpha
^{\prime }\beta ^{\prime }}=\delta _{\alpha ^{\prime }\beta ^{\prime }}.$ In
distinguished form, we obtain two skew matrices%
\begin{equation*}
\mathbf{\Gamma }_{h\mathbf{X}}^{i^{\prime }j^{\prime }}\doteqdot h\mathbf{%
X\rfloor \Gamma }_{\ }^{i^{\prime }j^{\prime }}=2\ e_{h\mathbf{X}%
}^{[i^{\prime }}\ \rho ^{j^{\prime }]}\mbox{ and }\mathbf{\Gamma }_{v\mathbf{%
X}}^{a^{\prime }b^{\prime }}\doteqdot v\mathbf{X\rfloor \Gamma }_{\
}^{a^{\prime }b^{\prime }}=2\mathbf{\ e}_{v\mathbf{X}}^{[a^{\prime }}\ \rho
^{b^{\prime }]}
\end{equation*}%
where
\begin{equation*}
\ e_{h\mathbf{X}}^{i^{\prime }}\doteqdot g(h\mathbf{X,}e^{i^{\prime }})=[1,%
\underbrace{0,\ldots ,0}_{n-1}]\mbox{ and }\ e_{v\mathbf{X}}^{a^{\prime
}}\doteqdot h(v\mathbf{X,}e^{a^{\prime }})=[1,\underbrace{0,\ldots ,0}_{m-1}]
\end{equation*}%
and
\begin{equation*}
\mathbf{\Gamma }_{h\mathbf{X\,}i^{\prime }}^{\qquad j^{\prime }}=\left[
\begin{array}{cc}
0 & \rho ^{j^{\prime }} \\
-\rho _{i^{\prime }} & \mathbf{0}_{[h]}%
\end{array}%
\right] \mbox{ and }\mathbf{\Gamma }_{v\mathbf{X\,}a^{\prime }}^{\qquad
b^{\prime }}=\left[
\begin{array}{cc}
0 & \rho ^{b^{\prime }} \\
-\rho _{a^{\prime }} & \mathbf{0}_{[v]}%
\end{array}%
\right]
\end{equation*}%
with $\mathbf{0}_{[h]}$ and $\mathbf{0}_{[v]}$ being respectively $%
(n-1)\times (n-1)$ and $(m-1)\times (m-1)$ matrices. The above presented
row--matrices and skew--matrices show that locally an N--anholonomic
manifold $\mathbf{V}$ of dimension $n+m,$ with respect to distinguished
orthonormalized frames are characterized algebraically by couples of unit
vectors in $\mathbb{R}^{n}$ and $\mathbb{R}^{m}$ preserved respectively by
the $SO(n-1)$ and $SO(m-1)$ rotation subgroups of the local N--adapted frame
structure group $SO(n)\oplus SO(m).$ The connection matrices $\mathbf{\Gamma
}_{h\mathbf{X\,}i^{\prime }}^{\qquad j^{\prime }}$ and $\mathbf{\Gamma }_{v%
\mathbf{X\,}a^{\prime }}^{\qquad b^{\prime }}$ belong to the orthogonal
complements of the corresponding Lie subalgebras and algebras, $\mathfrak{so}%
(n-1)\subset \mathfrak{so}(n)$ and $\mathfrak{so}(m-1)\subset \mathfrak{so}%
(m).$

The torsion (\ref{tors}) and curvature (\ref{curv}) (see Appendix) tensors
can be in orthonormalized component form with respect to (\ref{curvframe})
mapped into a distinguished orthotnomalized dual frame (\ref{orthbas}),%
\begin{equation}
\mathcal{T}^{\alpha ^{\prime }}\doteqdot \mathbf{D}_{\mathbf{X}}\mathbf{e}_{%
\mathbf{Y}}^{\alpha ^{\prime }}-\mathbf{D}_{\mathbf{Y}}\mathbf{e}_{\mathbf{X}%
}^{\alpha ^{\prime }}+\mathbf{e}_{\mathbf{Y}}^{\beta ^{\prime }}\Gamma _{%
\mathbf{X}\beta ^{\prime }}^{\quad \alpha ^{\prime }}-\mathbf{e}_{\mathbf{X}%
}^{\beta ^{\prime }}\Gamma _{\mathbf{Y}\beta ^{\prime }}^{\quad \alpha
^{\prime }}  \label{mtors}
\end{equation}%
and
\begin{equation}
\mathcal{R}_{\beta ^{\prime }}^{\;\alpha ^{\prime }}(\mathbf{X,Y})=\mathbf{D}%
_{\mathbf{Y}}\Gamma _{\mathbf{X}\beta ^{\prime }}^{\quad \alpha ^{\prime }}-%
\mathbf{D}_{\mathbf{X}}\Gamma _{\mathbf{Y}\beta ^{\prime }}^{\quad \alpha
^{\prime }}+\Gamma _{\mathbf{Y}\beta ^{\prime }}^{\quad \gamma ^{\prime
}}\Gamma _{\mathbf{X}\gamma ^{\prime }}^{\quad \alpha ^{\prime }}-\Gamma _{%
\mathbf{X}\beta ^{\prime }}^{\quad \gamma ^{\prime }}\Gamma _{\mathbf{Y}%
\gamma ^{\prime }}^{\quad \alpha ^{\prime }},  \label{mcurv}
\end{equation}%
where $\mathbf{e}_{\mathbf{Y}}^{\alpha ^{\prime }}\doteqdot \mathbf{g}(%
\mathbf{Y},\mathbf{e}^{\alpha ^{\prime }})$ and $\Gamma _{\mathbf{Y}\beta
^{\prime }}^{\quad \alpha ^{\prime }}\doteqdot \mathbf{Y\rfloor }\Gamma
_{\beta ^{\prime }}^{\;\alpha ^{\prime }}=\mathbf{g}(\mathbf{e}^{\alpha
^{\prime }},\mathbf{D}_{\mathbf{Y}}\mathbf{e}_{\beta ^{\prime }})$ define
respectively the N--adapted orthonormalized frame row--matrix and the
canonical d--connection skew--matrix in the flow directs, and
$\mathcal{R}_{\beta ^{\prime }}^{\;\alpha ^{\prime }} (\mathbf{X,Y})\doteqdot
\mathbf{g}(\mathbf{e}^{\alpha ^{\prime }}, [ \mathbf{D}_{\mathbf{X}},$ $
\mathbf{D}_{\mathbf{Y}}] \mathbf{e}_{\beta ^{\prime }})$
 is the curvature matrix. Both torsion and curvature components can be
distinguished in h-- and v--components like (\ref{dtors}) and (\ref{dcurv}),
by considering N--adapted decompositions of type
\begin{equation*}
\mathbf{g}=[g,h],\mathbf{e}_{\beta ^{\prime }}=(\mathbf{e}_{j^{\prime
}},e_{b^{\prime }}),\mathbf{e}^{\alpha ^{\prime }}=(e^{i^{\prime
}},e^{a^{\prime }}),\mathbf{X=}h\mathbf{X}+v\mathbf{X,D=(}h\mathbf{D},v%
\mathbf{D).}
\end{equation*}%
Finally, we note that the matrices for torsion (\ref{mtors}) and curvature (%
\ref{mcurv}) can be computed for any metric compatible linear connection
like the Levi Civita and the canonical d--connection. For our purposes, in
this work, we are interested to define such a frame of reference with
respect to which the curvature tensor has constant coefficients and the
torsion tensor vanishes.

\subsection{On anholonomic bundles with constant matrix curvature}

For vanishing N--connection curvature and torsion and constant matrix
curvature, we get a holonomic Riemannian manifold and the equations (\ref%
{mtors}) and (\ref{mcurv}) directly encode a bi--Hamiltonian structure, see
details in Refs. \cite{saw,anc2}. A well known class of Riemannian manifolds
for which the frame curvature matrix constant consists of the symmetric
spaces $M=G/H$ for compact semisimple Lie groups $G\supset H.$ A complete
classification and summary of main results on such spaces are given in Refs. %
\cite{helag,kob}. The Riemannian curvature and the metric tensors for $M=G/H$
are covariantly constant and $G$--invariant resulting in constant curvature
matrix. In \cite{anc1,anc2}, the bi--Hamiltonian operators were investigated
for the symmetric spaces with $M=G/SO(n)$ with $H=SO(n)\supset O(n-1)$ and
two examples when $G=SO(n+1),SU(n).$ Then it was exploited the existing
canonical soldering of Klein  and Riemannian symmetric--space
geometries \cite{sharpe}.

\subsubsection{Symmetric nonholonomic tangent bundles}

We suppose that the base manifold is a symmetric space $M=hG/SO(n)$ with the
isotropy subgroup $hH=SO(n)\supset O(n)$ and the typical fiber space to be a
symmetric space $F=vG/SO(m)$ with the isotropy subgroup $vH=SO(m)\supset
O(m).$ This means that $hG=SO(n+1)$ and $vG=SO(m+1)$ which is enough for a
study of real holonomic and nonholonomic manifolds and geometric mechanics
models.\footnote{%
it is necessary to consider $hG=SU(n)$ and $vG=SU(m)$ for the geometric
models with spinor and gauge fields} \

Our aim is to solder in a canonic way (like in the N--connection geometry)
the horizontal and vertical symmetric Riemannian spaces of dimension $n$ and
$m$ with a (total) symmetric Riemannian space $V$ of dimension $n+m,$ when $%
V=G/SO(n+m)$ with the isotropy group $H=SO(n+m)\supset O(n+m)$ and $%
G=SO(n+m+1).$ First, we note that for the just mentioned horizontal,
vertical and total symmetric Riemannian spaces one exists natural settings
to Klein geometry. For instance, the metric tensor $hg=\{\mathring{g}_{ij}\}$
on $M$ is defined by the Cartan--Killing inner product $<\cdot ,\cdot >_{h}$
on $T_{x}hG\simeq h\mathfrak{g}$ restricted to the Lie algebra quotient
spaces $h\mathfrak{p=}h\mathfrak{g/}h\mathfrak{h,}$ with $T_{x}hH\simeq h%
\mathfrak{h,}$ where $h\mathfrak{g=}h\mathfrak{h}\oplus h\mathfrak{p}$ is
stated such that there is an involutive automorphism of $hG$ under $hH$ is
fixed, i.e. $[h\mathfrak{h,}h\mathfrak{p]}\subseteq $ $h\mathfrak{p}$ and $[h%
\mathfrak{p,}h\mathfrak{p]}\subseteq h\mathfrak{h.}$ In a similar form, we
can define the group spaces and related inner products and\ Lie algebras,%
\begin{eqnarray}
\mbox{for\ }vg &=&\{\mathring{h}_{ab}\},\;<\cdot ,\cdot
>_{v},\;T_{y}vG\simeq v\mathfrak{g,\;}v\mathfrak{p=}v\mathfrak{g/}v\mathfrak{%
h,}\mbox{ with }  \notag \\
T_{y}vH &\simeq &v\mathfrak{h,}v\mathfrak{g=}v\mathfrak{h}\oplus v\mathfrak{%
p,}\mbox{where }\mathfrak{\;}[v\mathfrak{h,}v\mathfrak{p]}\subseteq v%
\mathfrak{p,\;}[v\mathfrak{p,}v\mathfrak{p]}\subseteq v\mathfrak{h;}  \notag
\\
&&  \label{algstr} \\
\mbox{for\ }\mathbf{g} &=&\{\mathring{g}_{\alpha \beta }\},\;<\cdot ,\cdot
>_{\mathbf{g}},\;T_{(x,y)}G\simeq \mathfrak{g,\;p=g/h,}\mbox{ with }  \notag
\\
T_{(x,y)}H &\simeq &\mathfrak{h,g=h}\oplus \mathfrak{p,}\mbox{where }%
\mathfrak{\;}[\mathfrak{h,p]}\subseteq \mathfrak{p,\;}[\mathfrak{p,p]}%
\subseteq \mathfrak{h.}  \notag
\end{eqnarray}%
We parametrize the metric structure with constant coefficients on $%
V=G/SO(n+m)$ in the form%
\begin{equation*}
\mathring{g}=\mathring{g}_{\alpha \beta }du^{\alpha }\otimes du^{\beta },
\end{equation*}%
where $u^{\alpha }$ are local coordinates and
\begin{equation}
\mathring{g}_{\alpha \beta }=\left[
\begin{array}{cc}
\mathring{g}_{ij}+\mathring{N}_{i}^{a}N_{j}^{b}\mathring{h}_{ab} & \mathring{%
N}_{j}^{e}\mathring{h}_{ae} \\
\mathring{N}_{i}^{e}\mathring{h}_{be} & \mathring{h}_{ab}%
\end{array}%
\right]  \label{constans}
\end{equation}%
when trivial, constant, N--connection coefficients are computed $\mathring{N}%
_{j}^{e}=\mathring{h}^{eb}\mathring{g}_{jb}$ for any given sets $\mathring{h}%
^{eb}$ and $\mathring{g}_{jb},$ i.e. from the inverse metrics coefficients
defined respectively on $hG=SO(n+1)$ and by off--blocks $(n\times n)$-- and $%
(m\times m)$--terms of the metric $\mathring{g}_{\alpha \beta }.$ As a
result, we define an equivalent d--metric structure of type (\ref{m1})
\begin{eqnarray}
\mathbf{\mathring{g}} &=&\ \mathring{g}_{ij}\ e^{i}\otimes e^{j}+\ \mathring{%
h}_{ab}\ \mathbf{\mathring{e}}^{a}\otimes \mathbf{\mathring{e}}^{b},
\label{m1const} \\
e^{i} &=&dx^{i},\ \;\mathbf{\mathring{e}}^{a}=dy^{a}+\mathring{N}%
_{i}^{e}dx^{i}  \notag
\end{eqnarray}%
defining a trivial $(n+m)$--splitting $\mathbf{\mathring{g}=}\mathring{g}%
\mathbf{\oplus _{\mathring{N}}}\mathring{h}\mathbf{\ }$because all
nonholonomy coefficients $\mathring{W}_{\alpha \beta }^{\gamma }$ and
N--connection curvature coefficients $\mathring{\Omega}_{ij}^{a}$ are zero.
In more general form, we can consider any covariant coordinate transforms of
(\ref{m1const}) preserving the\ $(n+m)$--splitting resulting in any $%
W_{\alpha \beta }^{\gamma }=0$ (\ref{anhrel}) and $\Omega _{ij}^{a}=0$ (\ref%
{ncurv}). It should be noted that even such trivial parametrizations define
algebraic classifications of \ symmetric Riemannian spaces of dimension $n+m$
with constant matrix curvature admitting splitting (by certain algebraic
constraints) into symmetric Riemannian subspaces of dimension $n$ and $m,$
also both with constant matrix curvature and introducing the concept of
N--anholonomic Riemannian space of type $\mathbf{\mathring{V}}=[hG=SO(n+1),$
$vG=SO(m+1),\;\mathring{N}_{i}^{e}].$ One can be considered that such
trivially N--anholonomic group spaces have possess a Lie d--algebra symmetry
$\mathfrak{so}_{\mathring{N}}(n+m)\doteqdot \mathfrak{so}(n)\oplus \mathfrak{%
so}(m).$

The simplest generalization on a vector bundle $\mathbf{\mathring{E}}$ is to
consider nonhlonomic distributions on $V=G/SO(n+m)$ defined locally by
arbitrary N--connection coefficients $N_{i}^{a}(x,y)$ with nonvanishing $%
W_{\alpha \beta }^{\gamma }$ and $\Omega _{ij}^{a}$ but with constant
d--metric coefficients when
\begin{eqnarray}
\mathbf{g} &=&\ \mathring{g}_{ij}\ e^{i}\otimes e^{j}+\ \mathring{h}_{ab}\
\mathbf{e}^{a}\otimes \mathbf{e}^{b},  \label{m1b} \\
e^{i} &=&dx^{i},\ \mathbf{e}^{a}=dy^{a}+N_{i}^{a}(x,y)dx^{i}.  \notag
\end{eqnarray}%
This metric is very similar to (\ref{clgs}) but with the coefficients $\
\mathring{g}_{ij}\ $\ and $\ \mathring{h}_{ab}$ induced by the corresponding
Lie d--algebra structure $\mathfrak{so}_{\mathring{N}}(n+m).$ Such spaces
transform into N--anholonomic Riemann--Cartan manifolds $\mathbf{\mathring{V}%
}_{\mathbf{N}}=[hG=SO(n+1),$ $vG=SO(m+1),\;N_{i}^{e}]$ with nontrivial
N--connection curvature and induced d--torsion coefficients of the canonical
d--connection (see formulas (\ref{dtors}) computed for constant d--metric
coefficients and the canonical d--connection coefficients in (\ref{candcon}%
)). One has zero curvature for the canonical d--connection (in general, such
spaces are curved ones with generic off--diagonal metric (\ref{m1b}) and
nonzero curvature tensor for the Levi Civita connection).\footnote{%
Introducing, constant values for the d--metric coefficients we get zero
coefficients for the canonical d--connection which in its turn results in
zero values of (\ref{dcurv}).} This allows us to classify the N--anholonomic
manifolds (and vector bundles) as having the same group and algebraic
structures of couples of symmetric Riemannian spaces of dimension $n$ and $m$
but nonholonomically soldered to the symmetric Riemannian space of dimension
$n+m.$ With respect to N--adapted orthonormal bases (\ref{orthbas}), with
distinguished h-- and v--subspaces, we obtain the same inner products and
group and Lie algebra spaces as in (\ref{algstr}).

The classification of N--anholonomic vector bundles is almost similar to
that for symmetric Riemannian spaces if we consider that $n=m$ and try to
model tangent bundles of such spaces, provided with N--connection structure.
For instance, we can take a (semi) Riemannian structure with the
N--connection induced by a absolute energy structure like in (\ref{cnlce})
and with the canonical d--connection structure (\ref{candcon}), for $\tilde{g%
}_{ef}=\ \mathring{g}_{ab},$ like in (\ref{aux4}). A straightforward
computation of the canonical d--connection coefficients\footnote{%
on tangent bundles, such d--connections can be defined to be torsionless}
and of d--curvatures for $\;^{\circ }\tilde{g}_{ij}$ and $\;^{\circ }%
\tilde{N}_{\ j}^{i}$ proves that the nonholonomic Riemanian manifold $\left(
M=SO(n+1)/SO(n),\;^{\circ }\mathcal{L}\right) $ possess constant both zero
canonical d--con\-nec\-ti\-on curvature and torsion but with induced
nontrivial N--connection curvature $\;^{\circ }\tilde{\Omega}_{jk}^{i}.$
Such spaces, being tangent to symmetric Riemannian spaces, are classified
similarly to the Riemannian ones with constant matrix curvature, see (\ref%
{algstr}) for $n=m$ but provided with a nonholonomic structure induced by
generating function $\;^{\circ }\mathcal{L}.$

\subsubsection{N--anholonomic Klein spaces}

The bi--Hamiltonian and solitonic constructions \cite{anc2,anc1,aw} are
based on an extrinsic approach soldering the Riemannian symmetric--space
geometry to the Klein geometry \cite{sharpe}. For the N--anhlonomic spaces
of dimension $n+n,$ with constant d--curvatures, similar constructions hold
true but we have to adapt them to the N--connection structure.

There are two Hamiltonian variables given by the principal normals $%
\;^{h}\nu $ and $\;^{v}\nu ,$ respectively, in the horizontal and vertical
subspaces, defined by the canonical d--connection $\mathbf{D}=(h\mathbf{D},v%
\mathbf{D}),$ see formulas (\ref{curvframe}) and (\ref{part02}),
\begin{equation*}
\;^{h}\nu \doteqdot \mathbf{D}_{h\mathbf{X}}h\mathbf{X}=\nu ^{\widehat{i}}%
\mathbf{\mathbf{e}}_{\widehat{i}}\mbox{\ and \ }\;^{v}\nu \doteqdot \mathbf{D%
}_{v\mathbf{X}}v\mathbf{X}=\nu ^{\widehat{a}}e_{\widehat{a}}.
\end{equation*}%
This normal d--vector $\mathbf{v}=(\;^{h}\nu ,$ $\;^{v}\nu ),$ with
components of type $\mathbf{\nu }^{\alpha }=(\nu ^{i},$ $\;\nu ^{a})=(\nu
^{1},$ $\nu ^{\widehat{i}},\nu ^{n+1},\nu ^{\widehat{a}}),$ is in the
tangent direction of curve $\gamma .$ There is also the principal normal
d--vector $\mathbf{\varpi }=(\;^{h}\varpi ,\;^{v}\varpi )$ with components
of type $\mathbf{\varpi }^{\alpha }=(\varpi ^{i},$ $\;\varpi ^{a})=(\varpi
^{1},\varpi ^{\widehat{i}},\varpi ^{n+1},\varpi ^{\widehat{a}})$ in the flow
direction, with
\begin{equation*}
\;^{h}\varpi \doteqdot \mathbf{D}_{h\mathbf{Y}}h\mathbf{X=}\varpi ^{\widehat{%
i}}\mathbf{\mathbf{e}}_{\widehat{i}},\;^{v}\varpi \doteqdot \mathbf{D}_{v%
\mathbf{Y}}v\mathbf{X}=\varpi ^{\widehat{a}}e_{\widehat{a}},
\end{equation*}%
representing a Hamiltonian d--covector field. We can consider that the normal
part of the flow d--vector
\begin{equation*}
\mathbf{h}_{\perp }\doteqdot \mathbf{Y}_{\perp }=h^{\widehat{i}}\mathbf{%
\mathbf{e}}_{\widehat{i}}+h^{\widehat{a}}e_{\widehat{a}}
\end{equation*}%
represents a Hamiltonian d--vector field. For such configurations, we can
consider parallel N--adapted frames $\mathbf{e}_{\alpha ^{\prime }}=(\mathbf{%
e}_{i^{\prime }},e_{a^{\prime }})$ when the h--variables $\nu ^{\widehat{%
i^{\prime }}},$ $\varpi ^{\widehat{i^{\prime }}},h^{\widehat{i^{\prime }}}$
are respectively encoded in the top row of the horizontal canonical
d--connection matrices $\mathbf{\Gamma }_{h\mathbf{X\,}i^{\prime }}^{\qquad
j^{\prime }}$ and $\mathbf{\Gamma }_{h\mathbf{Y\,}i^{\prime }}^{\qquad
j^{\prime }}$ and in the row matrix $\left( \mathbf{e}_{\mathbf{Y}%
}^{i^{\prime }}\right) _{\perp }\doteqdot \mathbf{e}_{\mathbf{Y}}^{i^{\prime
}}-g_{\parallel }\;\mathbf{e}_{\mathbf{X}}^{i^{\prime }}$ where $%
g_{\parallel }\doteqdot g(h\mathbf{Y,}h\mathbf{X})$ is the tangential
h--part of the flow d--vector. A similar encoding holds for v--variables $%
\nu ^{\widehat{a^{\prime }}},\varpi ^{\widehat{a^{\prime }}},h^{\widehat{%
a^{\prime }}}$ in the top row of the vertical canonical d--connection
matrices \ $\mathbf{\Gamma }_{v\mathbf{X\,}a^{\prime }}^{\qquad b^{\prime }}$
and $\mathbf{\Gamma }_{v\mathbf{Y\,}a^{\prime }}^{\qquad b^{\prime }}$ and
in the row matrix $\left( \mathbf{e}_{\mathbf{Y}}^{a^{\prime }}\right)
_{\perp }\doteqdot \mathbf{e}_{\mathbf{Y}}^{a^{\prime }}-h_{\parallel }\;%
\mathbf{e}_{\mathbf{X}}^{a^{\prime }}$ where $h_{\parallel }\doteqdot h(v%
\mathbf{Y,}v\mathbf{X})$ is the tangential v--part of the flow d--vector. In
a compact form of notations, we shall write $\mathbf{v}^{\alpha ^{\prime }}$
and $\mathbf{\varpi }^{\alpha ^{\prime }}$ where the primed small Greek
indices $\alpha ^{\prime },\beta ^{\prime },...$ will denote both N--adapted
and then orthonormalized components of geometric objects (like d--vectors,
d--covectors, d--tensors, d--groups, d--algebras, d--matrices) admitting
further decompositions into h-- and v--components defined as nonintegrable
distributions of such objects.

With respect to N--adapted orthonormalized frames, the geometry of
N--anholonomic manifolds is defined algebraically, on their tangent bundles,
by couples of horizontal and vertical Klein geometries considered in \cite%
{sharpe} and for bi--Hamiltonian soliton constructions in \cite{anc1}. The
N--connection structure induces a N--anholonomic Klein space stated by two
left--invariant $h\mathfrak{g}$-- and $v\mathfrak{g}$--valued Maurer--Cartan
form on the Lie d--group $\mathbf{G}=(h\mathbf{G},v\mathbf{G})$ is
identified with the zero--curvature canonical d--connection 1--form $\;^{%
\mathbf{G}}\mathbf{\Gamma }=\{\;^{\mathbf{G}}\mathbf{\Gamma }_{\ \beta
^{\prime }}^{\alpha ^{\prime }}\},$ where
\begin{equation*}
\;^{\mathbf{G}}\mathbf{\Gamma }_{\ \beta ^{\prime }}^{\alpha ^{\prime }}=\;^{%
\mathbf{G}}\mathbf{\Gamma }_{\ \beta ^{\prime }\gamma ^{\prime }}^{\alpha
^{\prime }}\mathbf{e}^{\gamma ^{\prime }}=\;^{h\mathbf{G}}L_{\;j^{\prime
}k^{\prime }}^{i^{\prime }}\mathbf{e}^{k^{\prime }}+\;^{v\mathbf{G}%
}C_{\;j^{\prime }k^{\prime }}^{i^{\prime }}e^{k^{\prime }}.
\end{equation*}%
For trivial N--connection structure in vector bundles with the base and
typical fiber spaces being symmetric Riemannian spaces, we can consider that
$\;^{h\mathbf{G}}L_{\;j^{\prime }k^{\prime }}^{i^{\prime }}$ and $\;^{v%
\mathbf{G}}C_{\;j^{\prime }k^{\prime }}^{i^{\prime }}$ are the coefficients
of the Cartan connections $\;^{h\mathbf{G}}L$ and $\;^{v\mathbf{G}}C,$
respectively for the $h\mathbf{G}$ and $v\mathbf{G,}$ both with vanishing
curvatures, i.e. with
\begin{equation*}
d\;^{\mathbf{G}}\mathbf{\Gamma +}\frac{1}{2}\mathbf{[\;^{\mathbf{G}}\mathbf{%
\Gamma ,}\;^{\mathbf{G}}\mathbf{\Gamma }]=0}
\end{equation*}%
and h-- and v--components,
$d\;^{h\mathbf{G}}\mathbf{L}+\frac{1}{2}\mathbf{[\;^{h\mathbf{G}}L\%
\;^{h\mathbf{G}}L]}=0$ and $d\;^{v\mathbf{G}}\mathbf{C}+\frac{1}{2}%
[\;^{v\mathbf{G}}\mathbf{C},$ $ \;^{v\mathbf{G}}\mathbf{C}]=0, $
where $d$ denotes the total derivatives on the d--group manifold $\mathbf{G}%
=h\mathbf{G}\oplus v\mathbf{G}$ or their restrictions on $h\mathbf{G}$ or $v%
\mathbf{G.}$ We can consider that $\;^{\mathbf{G}}\mathbf{\Gamma }$ defines
the so--called Cartan d--connection for nonintegrable N--connection
structures, see details and supersymmetric/ noncommutative developments in %
\cite{vncg,vsgg}.

Through the Lie d--algebra decompositions $\mathfrak{g}=h\mathfrak{g}\oplus v%
\mathfrak{g,}$ for the horizontal splitting: $h\mathfrak{g}=\mathfrak{so}%
(n)\oplus h\mathfrak{p,}$ when $[h\mathfrak{p},h\mathfrak{p}]\subset
\mathfrak{so}(n)$ and $[\mathfrak{so}(n),h\mathfrak{p}]\subset h\mathfrak{p;}
$ for the vertical splitting $v\mathfrak{g}=\mathfrak{so}(m)\oplus v%
\mathfrak{p,}$ when $[v\mathfrak{p},v\mathfrak{p}]\subset \mathfrak{so}(m)$
and $[\mathfrak{so}(m),v\mathfrak{p}]\subset v\mathfrak{p,}$ the Cartan
d--connection determines an N--anholonomic Riemannian structure on the
nonholonomic bundle $\mathbf{\mathring{E}}=[hG=SO(n+1),$ $%
vG=SO(m+1),\;N_{i}^{e}].$ For $n=m,$ and canonical d--objects
(N--connection, d--metric, d--connection, ...) derived from (\ref{m1b}), or
any N--anholonomic space with constant d--curvatures, the Cartan
d--connection transform just in the canonical d--connection (\ref{candcontm}%
). It is possible to consider a quotient space with distinguished structure
group $\mathbf{V}_{\mathbf{N}}=\mathbf{G}/SO(n)\oplus $ $SO(m)$ regarding $%
\mathbf{G}$ as a principal $\left( SO(n)\oplus SO(m)\right) $--bundle over $%
\mathbf{\mathring{E}},$ which is a N--anholonomic bundle. In this case, we
can always fix a local section of this bundle and pull--back $\;^{\mathbf{G}}%
\mathbf{\Gamma }$ to give a $\left( h\mathfrak{g}\oplus v\mathfrak{g}\right)
$--valued 1--form $^{\mathfrak{g}}\mathbf{\Gamma }$ in a point $u\in \mathbf{%
\mathring{E}}.$ Any change of local sections define $SO(n)\oplus $ $SO(m)$
gauge transforms of the canonical d--connection $^{\mathfrak{g}}\mathbf{%
\Gamma ,}$ all preserving the nonholonomic decomposition (\ref{whitney}).

There are involutive automorphisms $h\sigma =\pm 1$ and $v\sigma =\pm 1,$
respectively, of $h\mathfrak{g}$ and $v\mathfrak{g,}$ defined that $%
\mathfrak{so}(n)$ (or $\mathfrak{so}(m)$) is eigenspace $h\sigma =+1$ (or $%
v\sigma =+1)$ and $h\mathfrak{p}$ (or $v\mathfrak{p}$) is eigenspace $%
h\sigma =-1$ (or $v\sigma =-1).$ It is possible both a N--adapted
decomposition and taking into account the existing eigenspaces, when the
symmetric parts%
\begin{equation*}
\mathbf{\Gamma \doteqdot }\frac{1}{2}\left( ^{\mathfrak{g}}\mathbf{\Gamma +}%
\sigma \left( ^{\mathfrak{g}}\mathbf{\Gamma }\right) \right) ,
\end{equation*}%
with respective h- and v--splitting
$\mathbf{L\doteqdot }\frac{1}{2}\left( ^{h\mathfrak{g}}\mathbf{L+}h\sigma
\left( ^{h\mathfrak{g}}\mathbf{L}\right) \right)$  and
$\mathbf{C\doteqdot }\frac{1}{2}( ^{v\mathfrak{g}}\mathbf{C}
+h\sigma ( ^{v \mathfrak{g}}\mathbf{C})),$
 defines a $\left( \mathfrak{so}(n)\oplus \mathfrak{so}(m)\right) $--valued
d--connection 1--form. Under such conditions, the antisymmetric part
\begin{equation*}
\mathbf{e\doteqdot }\frac{1}{2}\left( ^{\mathfrak{g}}\mathbf{\Gamma -}\sigma
\left( ^{\mathfrak{g}}\mathbf{\Gamma }\right) \right) ,
\end{equation*}%
with respective h- and v--splitting
$ h\mathbf{e\doteqdot }\frac{1}{2}\left( ^{h\mathfrak{g}}\mathbf{L-}h\sigma
\left( ^{h\mathfrak{g}}\mathbf{L}\right) \right)$  and
$v\mathbf{%
e\doteqdot }\frac{1}{2} ( ^{v\mathfrak{g}}\mathbf{C} - h\sigma ( ^{v%
\mathfrak{g}}\mathbf{C})),$
 defines a $\left( h\mathfrak{p}\oplus v\mathfrak{p}\right) $--valued
N--adapted coframe for the Cartan--Killing inner product $<\cdot ,\cdot >_{%
\mathfrak{p}}$ on $T_{u}\mathbf{G}\simeq h\mathfrak{g}\oplus v\mathfrak{g}$
restricted to $T_{u}\mathbf{V}_{\mathbf{N}}\simeq \mathfrak{p.}$ This inner
product, distinguished into h- and v--components, provides a d--metric
structure of type $\mathbf{g}=[g,h]$ (\ref{m1}),where
 $g=<h\mathbf{e\otimes }h\mathbf{e}>_{h\mathfrak{p}}$ and $h=<v\mathbf{e\otimes }v%
\mathbf{e}>_{v\mathfrak{p}}$
on $\mathbf{V}_{\mathbf{N}}=\mathbf{G}/SO(n)\oplus $ $SO(m).$

We generate a $\mathbf{G(}=h\mathbf{G}\oplus v\mathbf{G)}$--invariant
d--derivative $\mathbf{D}$ whose restriction to the tangent space $T\mathbf{V%
}_{\mathbf{N}}$ for any N--anholonomic curve flow $\gamma (\tau ,\mathbf{l})$
in $\mathbf{V}_{\mathbf{N}}=\mathbf{G}/SO(n)\oplus $ $SO(m)$ is defined via%
\begin{equation}
\mathbf{D}_{\mathbf{X}}\mathbf{e=}\left[ \mathbf{e},\gamma _{\mathbf{l}%
}\rfloor \mathbf{\Gamma }\right] \mbox{\ and \ }\mathbf{D}_{\mathbf{Y}}%
\mathbf{e=}\left[ \mathbf{e},\gamma _{\mathbf{\tau }}\rfloor \mathbf{\Gamma }%
\right] ,  \label{aux33}
\end{equation}%
admitting further h- and v--decompositions. The derivatives $\mathbf{D}_{%
\mathbf{X}}$ and $\mathbf{D}_{\mathbf{Y}}$ are equivalent to those
considered in (\ref{part01}) and obey the Cartan structure equations (\ref%
{mtors}) and (\ref{mcurv}). For the canonical d--connections, a large class
of N--anholonomic spaces of dimension $n=m,$ the d--torsions are zero and
the d--curvatures are with constant coefficients.

Let $\mathbf{e}^{\alpha ^{\prime }}=(e^{i^{\prime }},\mathbf{e}^{a^{\prime
}})$ be a N--adapted orthonormalized coframe being identified with the $%
\left( h\mathfrak{p}\oplus v\mathfrak{p}\right) $--valued coframe $\mathbf{e}
$ in a fixed orthonormal basis for $\mathfrak{p=}h\mathfrak{p}\oplus v%
\mathfrak{p\subset }h\mathfrak{g}\oplus v\mathfrak{g.}$ Considering the
kernel/ cokernel of Lie algebra multiplications in the h- and v--subspaces,
respectively, $\left[ \mathbf{e}_{h\mathbf{X}},\cdot \right] _{h\mathfrak{g}%
} $ and $\left[ \mathbf{e}_{v\mathbf{X}},\cdot \right] _{v\mathfrak{g}},$ we
can decompose the coframes into parallel and perpendicular parts with
respect to $\mathbf{e}_{\mathbf{X}}.$ We write
\begin{equation*}
\mathbf{e=(e}_{C}=h\mathbf{e}_{C}+v\mathbf{e}_{C},\mathbf{e}_{C^{\perp }}=h%
\mathbf{e}_{C^{\perp }}+v\mathbf{e}_{C^{\perp }}\mathbf{),}
\end{equation*}%
for $\mathfrak{p(}=h\mathfrak{p}\oplus v\mathfrak{p)}$--valued mutually
orthogonal d--vectors $\mathbf{e}_{C}$ \ and $\mathbf{e}_{C^{\perp }},$ when
there are satisfied the conditions $\left[ \mathbf{e}_{\mathbf{X}},\mathbf{e}%
_{C}\right] _{\mathfrak{g}}=0$ but $\left[ \mathbf{e}_{\mathbf{X}},\mathbf{e}%
_{C^{\perp }}\right] _{\mathfrak{g}}\neq 0;$ such conditions can be stated
in h- and v--component form, respectively, $\left[ h\mathbf{e}_{\mathbf{X}},h%
\mathbf{e}_{C}\right] _{h\mathfrak{g}}=0,$ $\left[ h\mathbf{e}_{\mathbf{X}},h%
\mathbf{e}_{C^{\perp }}\right] _{h\mathfrak{g}}\neq 0$ and $\left[ v\mathbf{e%
}_{\mathbf{X}},v\mathbf{e}_{C}\right] _{v\mathfrak{g}}=0,$ $\left[ v\mathbf{e%
}_{\mathbf{X}},v\mathbf{e}_{C^{\perp }}\right] _{v\mathfrak{g}}\neq 0.$ One
holds also the algebraic decompositions
\begin{equation*}
T_{u}\mathbf{V}_{\mathbf{N}}\simeq \mathfrak{p=}h\mathfrak{p}\oplus v%
\mathfrak{p}=\mathfrak{g=}h\mathfrak{g}\oplus v\mathfrak{g}/\mathfrak{so}%
(n)\oplus \mathfrak{so}(m)
\end{equation*}%
and
\begin{equation*}
\mathfrak{p=p}_{C}\oplus \mathfrak{p}_{C^{\perp }}=\left( h\mathfrak{p}%
_{C}\oplus v\mathfrak{p}_{C}\right) \oplus \left( h\mathfrak{p}_{C^{\perp
}}\oplus v\mathfrak{p}_{C^{\perp }}\right) ,
\end{equation*}%
with $\mathfrak{p}_{\parallel }\subseteq \mathfrak{p}_{C}$ and $\mathfrak{p}%
_{C^{\perp }}\subseteq \mathfrak{p}_{\perp },$ where $\left[ \mathfrak{p}%
_{\parallel },\mathfrak{p}_{C}\right] =0,$ $<\mathfrak{p}_{C^{\perp }},%
\mathfrak{p}_{C}>=0,$ but $\left[ \mathfrak{p}_{\parallel },\mathfrak{p}%
_{C^{\perp }}\right] \neq 0$ (i.e. $\mathfrak{p}_{C}$ is the centralizer of $%
\mathbf{e}_{\mathbf{X}}$ in $\mathfrak{p=}h\mathfrak{p}\oplus v\mathfrak{%
p\subset }h\mathfrak{g}\oplus v\mathfrak{g);}$ in h- \ and v--components,
one have $h\mathfrak{p}_{\parallel }\subseteq h\mathfrak{p}_{C}$ and $h%
\mathfrak{p}_{C^{\perp }}\subseteq h\mathfrak{p}_{\perp },$ where $\left[ h%
\mathfrak{p}_{\parallel },h\mathfrak{p}_{C}\right] =0,$ $<h\mathfrak{p}%
_{C^{\perp }},h\mathfrak{p}_{C}>=0,$ but $\left[ h\mathfrak{p}_{\parallel },h%
\mathfrak{p}_{C^{\perp }}\right] \neq 0$ (i.e. $h\mathfrak{p}_{C}$ is the
centralizer of $\mathbf{e}_{h\mathbf{X}}$ in $h\mathfrak{p\subset }h%
\mathfrak{g)}$ and $v\mathfrak{p}_{\parallel }\subseteq v\mathfrak{p}_{C}$
and $v\mathfrak{p}_{C^{\perp }}\subseteq v\mathfrak{p}_{\perp },$ where $%
\left[ v\mathfrak{p}_{\parallel },v\mathfrak{p}_{C}\right] =0,$ $<v\mathfrak{%
p}_{C^{\perp }},v\mathfrak{p}_{C}>=0,$ but $\left[ v\mathfrak{p}_{\parallel
},v\mathfrak{p}_{C^{\perp }}\right] \neq 0$ (i.e. $v\mathfrak{p}_{C}$ is the
centralizer of $\mathbf{e}_{v\mathbf{X}}$ in $v\mathfrak{p\subset }v%
\mathfrak{g).}$ Using the canonical d--connection derivative $\mathbf{D}_{%
\mathbf{X}}$ of a d--covector perpendicular (or parallel) to $\mathbf{e}_{%
\mathbf{X}},$ we get a new d--vector which is parallel (or perpendicular) to
$\mathbf{e}_{\mathbf{X}},$ i.e. $\mathbf{D}_{\mathbf{X}}\mathbf{e}_{C}\in
\mathfrak{p}_{C^{\perp }}$ (or $\mathbf{D}_{\mathbf{X}}\mathbf{e}_{C^{\perp
}}\in \mathfrak{p}_{C});$ in h- \ and v--components such formulas are
written $\mathbf{D}_{h\mathbf{X}}h\mathbf{e}_{C}\in h\mathfrak{p}_{C^{\perp
}}$ (or $\mathbf{D}_{h\mathbf{X}}h\mathbf{e}_{C^{\perp }}\in h\mathfrak{p}%
_{C})$ and $\mathbf{D}_{v\mathbf{X}}v\mathbf{e}_{C}\in v\mathfrak{p}%
_{C^{\perp }}$ (or $\mathbf{D}_{v\mathbf{X}}v\mathbf{e}_{C^{\perp }}\in v%
\mathfrak{p}_{C}).$ All such d--algebraic relations can be written in
N--anholonomic manifolds and canonical d--connection settings, for instance,
using certain relations of type
\begin{equation*}
\mathbf{D}_{\mathbf{X}}(\mathbf{e}^{\alpha ^{\prime }})_{C}=\mathbf{v}%
_{~\beta ^{\prime }}^{\alpha ^{\prime }}(\mathbf{e}^{\beta ^{\prime
}})_{C^{\perp }}\mbox{ \ and \ }\mathbf{D}_{\mathbf{X}}(\mathbf{e}^{\alpha
^{\prime }})_{C^{\perp }}=-\mathbf{v}_{~\beta ^{\prime }}^{\alpha ^{\prime
}}(\mathbf{e}^{\beta ^{\prime }})_{C},
\end{equation*}%
for some antisymmetric d--tensors $\mathbf{v}^{\alpha ^{\prime }\beta
^{\prime }}=-\mathbf{v}^{\beta ^{\prime }\alpha ^{\prime }}.$ We get a
N--adapted $\left( SO(n)\oplus SO(m)\right) $--parallel frame defining a
generalization of the concept of Riemannian parallel frame on N--adapted
manifolds whenever $\mathfrak{p}_{C}$ is larger than $\mathfrak{p}%
_{\parallel }.$ Substituting $\mathbf{e}^{\alpha ^{\prime }}=(e^{i^{\prime
}},\mathbf{e}^{a^{\prime }})$ into the last formulas and considering h- and
v--components, we define $SO(n)$--parallel and $SO(m)$--parallel frames (for
simplicity we omit these formulas when the Greek small letter indices are
split into Latin small letter h- and v--indices).

The final conclusion of this section is that the Cartan structure equations on
hypersurfaces swept out by nonholonomic curve flows on N--anholonomic spaces
with constant matrix curvature for the canonical d--connection geometrically
encode \ two $O(n-1)$-- and $O(m-1)$--invariant, respectively, horizontal
and vertical bi--Hamiltonian operators. This holds true if the distinguished
by N--connection freedom of the d--group action $SO(n)\oplus SO(m)$ on $%
\mathbf{e}$ and $\mathbf{\Gamma }$ is used to fix them to be a N--adapted
parallel coframe and its associated canonical d--connection 1--form is
related to the canonical covariant derivative on N--anholonomic manifolds.

\section{Anholonomic bi--Hamiltonians and Vector Solitons}

Introducing N--adapted orthonormalized bases, for N--anholonomic spaces of
dimension $n+n,$ with constant curvatures of the canonical d--connection,
we can derive bi--Hamiltonian and vector soliton structures similarly to %
\cite{anc2,anc1,aw}. In symbolic, abstract index form, the constructions for
nonholonomic vector bundles are similar to those for the Riemannian
symmetric--spaces soldered to Klein geometry. We have to distinguish the
horizontal and vertical components of geometric objects and related
equations.

\subsection{Basic equations for N--anholonomic curve flows}

In this section, we shall prove the results for the h--components of certain
N--anholonomic manifolds with constant d--curvature and then dub the
formulas for the v--components omitting similar details.

There is an isomorphism between the real space $\mathfrak{so}(n)$ and the
Lie algebra of $n\times n$ skew--symmetric matrices. This allows to
establish an isomorphism between $h\mathfrak{p}$ $\simeq \mathbb{R}^{n}$ and
the tangent spaces $T_{x}M=\mathfrak{so}(n+1)/$ $\mathfrak{so}(n)$ of the
Riemannian manifold $M=SO(n+1)/$ $SO(n)$ as described by the following
canonical decomposition
\begin{equation*}
h\mathfrak{g}=\mathfrak{so}(n+1)\supset h\mathfrak{p\in }\left[
\begin{array}{cc}
0 & h\mathbf{p} \\
-h\mathbf{p}^{T} & h\mathbf{0}%
\end{array}%
\right] \mbox{\ for\ }h\mathbf{0\in }h\mathfrak{h=so}(n)
\end{equation*}%
with $h\mathbf{p=\{}p^{i^{\prime }}\mathbf{\}\in }\mathbb{R}^{n}$ being the
h--component of the d--vector $\mathbf{p=(}p^{i^{\prime }}\mathbf{,}%
p^{a^{\prime }}\mathbf{)}$ and $h\mathbf{p}^{T}$ mean the transposition of
the row $h\mathbf{p.}$ The Cartan--Killing inner product on $h\mathfrak{g}$
is stated following the rule%
\begin{eqnarray*}
h\mathbf{p\cdot }h\mathbf{p} &\mathbf{=}&\left\langle \left[
\begin{array}{cc}
0 & h\mathbf{p} \\
-h\mathbf{p}^{T} & h\mathbf{0}%
\end{array}%
\right] ,\left[
\begin{array}{cc}
0 & h\mathbf{p} \\
-h\mathbf{p}^{T} & h\mathbf{0}%
\end{array}%
\right] \right\rangle \\
&\mathbf{\doteqdot }&\frac{1}{2}tr\left\{ \left[
\begin{array}{cc}
0 & h\mathbf{p} \\
-h\mathbf{p}^{T} & h\mathbf{0}%
\end{array}%
\right] ^{T}\left[
\begin{array}{cc}
0 & h\mathbf{p} \\
-h\mathbf{p}^{T} & h\mathbf{0}%
\end{array}%
\right] \right\} ,
\end{eqnarray*}%
where $tr$ denotes the trace of the corresponding product of matrices. This
product identifies canonically $h\mathfrak{p}$ $\simeq \mathbb{R}^{n}$ with
its dual $h\mathfrak{p}^{\ast }$ $\simeq \mathbb{R}^{n}.$ In a similar form,
we can consider
\begin{equation*}
v\mathfrak{g}=\mathfrak{so}(m+1)\supset v\mathfrak{p\in }\left[
\begin{array}{cc}
0 & v\mathbf{p} \\
-v\mathbf{p}^{T} & v\mathbf{0}%
\end{array}%
\right] \mbox{\ for\ }v\mathbf{0\in }v\mathfrak{h=so}(m)
\end{equation*}%
with $v\mathbf{p=\{}p^{a^{\prime }}\mathbf{\}\in }\mathbb{R}^{m}$ being the
v--component of the d--vector $\mathbf{p=(}p^{i^{\prime }}\mathbf{,}%
p^{a^{\prime }}\mathbf{)}$ and define the Cartan--Killing inner product $v%
\mathbf{p\cdot }v\mathbf{p\doteqdot }\frac{1}{2}tr\{...\}.$ In general, in
the tangent bundle of a N--anholonomic manifold, we can consider the
Cartan--Killing N--adapted inner product $\mathbf{p\cdot p=}h\mathbf{p\cdot }%
h\mathbf{p+}v\mathbf{p\cdot }v\mathbf{p.}$

Following the introduced Cartan--Killing parametrizations, we analyze the
flow $\gamma (\tau ,\mathbf{l})$ of a non--stretching curve in $\mathbf{V}_{%
\mathbf{N}}=\mathbf{G}/SO(n)\oplus $ $SO(m).$ Let us introduce a coframe $%
\mathbf{e}\in T_{\gamma }^{\ast }\mathbf{V}_{\mathbf{N}}\otimes (h\mathfrak{%
p\oplus }v\mathfrak{p}),$ which is a N--adapted $\left( SO(n)\mathfrak{%
\oplus }SO(m)\right) $--parallel basis along $\gamma ,$ and its associated
canonical d--con\-nec\-tion 1--form $\mathbf{\Gamma }\in T_{\gamma }^{\ast }%
\mathbf{V}_{\mathbf{N}}\otimes (\mathfrak{so}(n)\mathfrak{\oplus so}(m)).$
Such d--objects are respectively parametrized:%
\begin{equation*}
\mathbf{e}_{\mathbf{X}}=\mathbf{e}_{h\mathbf{X}}+\mathbf{e}_{v\mathbf{X}},
\end{equation*}%
for
\begin{equation*}
\mathbf{e}_{h\mathbf{X}}=\gamma _{h\mathbf{X}}\rfloor h\mathbf{e=}\left[
\begin{array}{cc}
0 & (1,\overrightarrow{0}) \\
-(1,\overrightarrow{0})^{T} & h\mathbf{0}%
\end{array}%
\right]
\end{equation*}%
and
\begin{equation*}
\mathbf{e}_{v\mathbf{X}}=\gamma _{v\mathbf{X}}\rfloor v\mathbf{e=}\left[
\begin{array}{cc}
0 & (1,\overleftarrow{0}) \\
-(1,\overleftarrow{0})^{T} & v\mathbf{0}%
\end{array}%
\right] ,
\end{equation*}%
where we write $(1,\overrightarrow{0})\in \mathbb{R}^{n},\overrightarrow{0}%
\in \mathbb{R}^{n-1}$ and $(1,\overleftarrow{0})\in \mathbb{R}^{m},%
\overleftarrow{0}\in \mathbb{R}^{m-1};$%
\begin{equation*}
\mathbf{\Gamma =}\left[ \mathbf{\Gamma }_{h\mathbf{X}},\mathbf{\Gamma }_{v%
\mathbf{X}}\right] ,
\end{equation*}%
for
\begin{equation*}
\mathbf{\Gamma }_{h\mathbf{X}}\mathbf{=}\gamma _{h\mathbf{X}}\rfloor \mathbf{%
L=}\left[
\begin{array}{cc}
0 & (0,\overrightarrow{0}) \\
-(0,\overrightarrow{0})^{T} & \mathbf{L}%
\end{array}%
\right] \in \mathfrak{so}(n+1),
\end{equation*}%
where
\begin{equation*}
\mathbf{L=}\left[
\begin{array}{cc}
0 & \overrightarrow{v} \\
-\overrightarrow{v}^{T} & h\mathbf{0}%
\end{array}%
\right] \in \mathfrak{so}(n),~\overrightarrow{v}\in \mathbb{R}^{n-1},~h%
\mathbf{0\in }\mathfrak{so}(n-1),
\end{equation*}%
and
\begin{equation*}
\mathbf{\Gamma }_{v\mathbf{X}}\mathbf{=}\gamma _{v\mathbf{X}}\rfloor \mathbf{%
C=}\left[
\begin{array}{cc}
0 & (0,\overleftarrow{0}) \\
-(0,\overleftarrow{0})^{T} & \mathbf{C}%
\end{array}%
\right] \in \mathfrak{so}(m+1),
\end{equation*}%
where
\begin{equation*}
\mathbf{C=}\left[
\begin{array}{cc}
0 & \overleftarrow{v} \\
-\overleftarrow{v}^{T} & v\mathbf{0}%
\end{array}%
\right] \in \mathfrak{so}(m),~\overleftarrow{v}\in \mathbb{R}^{m-1},~v%
\mathbf{0\in }\mathfrak{so}(m-1).
\end{equation*}%
The above parametrizations are fixed in order to preserve the $SO(n)$ and $%
SO(m)$ rotation gauge freedoms on the N--adapted coframe and canonical
d--connec\-ti\-on 1--form, distinguished in h- and v--components.

There are defined decompositions of horizontal $SO(n+1)/$ $SO(n)$ matrices
like%
\begin{eqnarray*}
h\mathfrak{p} &\mathfrak{\ni }&\left[
\begin{array}{cc}
0 & h\mathbf{p} \\
-h\mathbf{p}^{T} & h\mathbf{0}%
\end{array}%
\right] =\left[
\begin{array}{cc}
0 & \left( h\mathbf{p}_{\parallel },\overrightarrow{0}\right) \\
-\left( h\mathbf{p}_{\parallel },\overrightarrow{0}\right) ^{T} & h\mathbf{0}%
\end{array}%
\right] \\
&&+\left[
\begin{array}{cc}
0 & \left( 0,h\overrightarrow{\mathbf{p}}_{\perp }\right) \\
-\left( 0,h\overrightarrow{\mathbf{p}}_{\perp }\right) ^{T} & h\mathbf{0}%
\end{array}%
\right] ,
\end{eqnarray*}%
into tangential and normal parts relative to $\mathbf{e}_{h\mathbf{X}}$ via
corresponding decompositions of h--vectors $h\mathbf{p=(}h\mathbf{\mathbf{p}%
_{\parallel },}h\mathbf{\overrightarrow{\mathbf{p}}_{\perp })\in }\mathbb{R}%
^{n}$ relative to $\left( 1,\overrightarrow{0}\right) ,$ when $h\mathbf{%
\mathbf{p}_{\parallel }}$ is identified with $h\mathfrak{p}_{C}$ and $h%
\mathbf{\overrightarrow{\mathbf{p}}_{\perp }}$ is identified with $h%
\mathfrak{p}_{\perp }=h\mathfrak{p}_{C^{\perp }}.$ In a similar form, it is
possible to decompose vertical $SO(m+1)/$ $SO(m)$ matrices,
\begin{eqnarray*}
v\mathfrak{p} &\mathfrak{\ni }&\left[
\begin{array}{cc}
0 & v\mathbf{p} \\
-v\mathbf{p}^{T} & v\mathbf{0}%
\end{array}%
\right] =\left[
\begin{array}{cc}
0 & \left( v\mathbf{p}_{\parallel },\overleftarrow{0}\right) \\
-\left( v\mathbf{p}_{\parallel },\overleftarrow{0}\right) ^{T} & v\mathbf{0}%
\end{array}%
\right] \\
&&+\left[
\begin{array}{cc}
0 & \left( 0,v\overleftarrow{\mathbf{p}}_{\perp }\right) \\
-\left( 0,v\overleftarrow{\mathbf{p}}_{\perp }\right) ^{T} & v\mathbf{0}%
\end{array}%
\right] ,
\end{eqnarray*}%
into tangential and normal parts relative to $\mathbf{e}_{v\mathbf{X}}$ via
corresponding decompositions of h--vectors $v\mathbf{p=(}v\mathbf{\mathbf{p}%
_{\parallel },}v\overleftarrow{\mathbf{\mathbf{p}}}\mathbf{_{\perp })\in }%
\mathbb{R}^{m}$ relative to $\left( 1,\overleftarrow{0}\right) ,$ when $v%
\mathbf{\mathbf{p}_{\parallel }}$ is identified with $v\mathfrak{p}_{C}$ and
$v\overleftarrow{\mathbf{\mathbf{p}}}\mathbf{_{\perp }}$ is identified with $%
v\mathfrak{p}_{\perp }=v\mathfrak{p}_{C^{\perp }}.$

The canonical d--connection induces matrices decomposed with respect to the
flow direction. In the h--direction, we parametrize%
\begin{equation*}
\mathbf{e}_{h\mathbf{Y}}=\gamma _{\tau }\rfloor h\mathbf{e=}\left[
\begin{array}{cc}
0 & \left( h\mathbf{e}_{\parallel },h\overrightarrow{\mathbf{e}}_{\perp
}\right) \\
-\left( h\mathbf{e}_{\parallel },h\overrightarrow{\mathbf{e}}_{\perp
}\right) ^{T} & h\mathbf{0}%
\end{array}%
\right] ,
\end{equation*}%
when $\mathbf{e}_{h\mathbf{Y}}\in h\mathfrak{p,}\left( h\mathbf{e}%
_{\parallel },h\overrightarrow{\mathbf{e}}_{\perp }\right) \in \mathbb{R}%
^{n} $ and $h\overrightarrow{\mathbf{e}}_{\perp }\in \mathbb{R}^{n-1},$ and
\begin{equation}
\mathbf{\Gamma }_{h\mathbf{Y}}\mathbf{=}\gamma _{h\mathbf{Y}}\rfloor \mathbf{%
L=}\left[
\begin{array}{cc}
0 & (0,\overrightarrow{0}) \\
-(0,\overrightarrow{0})^{T} & h\mathbf{\varpi }_{\tau }%
\end{array}%
\right] \in \mathfrak{so}(n+1),  \label{auxaaa}
\end{equation}%
where
\begin{equation*}
h\mathbf{\varpi }_{\tau }\mathbf{=}\left[
\begin{array}{cc}
0 & \overrightarrow{\varpi } \\
-\overrightarrow{\varpi }^{T} & h\mathbf{\Theta }%
\end{array}%
\right] \in \mathfrak{so}(n),~\overrightarrow{\varpi }\in \mathbb{R}^{n-1},~h%
\mathbf{\Theta \in }\mathfrak{so}(n-1).
\end{equation*}%
In the v--direction, we parametrize
\begin{equation*}
\mathbf{e}_{v\mathbf{Y}}=\gamma _{\tau }\rfloor v\mathbf{e=}\left[
\begin{array}{cc}
0 & \left( v\mathbf{e}_{\parallel },v\overleftarrow{\mathbf{e}}_{\perp
}\right) \\
-\left( v\mathbf{e}_{\parallel },v\overleftarrow{\mathbf{e}}_{\perp }\right)
^{T} & v\mathbf{0}%
\end{array}%
\right] ,
\end{equation*}%
when $\mathbf{e}_{v\mathbf{Y}}\in v\mathfrak{p,}\left( v\mathbf{e}%
_{\parallel },v\overleftarrow{\mathbf{e}}_{\perp }\right) \in \mathbb{R}^{m}$
and $v\overleftarrow{\mathbf{e}}_{\perp }\in \mathbb{R}^{m-1},$ and
\begin{equation*}
\mathbf{\Gamma }_{v\mathbf{Y}}\mathbf{=}\gamma _{v\mathbf{Y}}\rfloor \mathbf{%
C=}\left[
\begin{array}{cc}
0 & (0,\overleftarrow{0}) \\
-(0,\overleftarrow{0})^{T} & v\mathbf{\varpi }_{\tau }%
\end{array}%
\right] \in \mathfrak{so}(m+1),
\end{equation*}%
where
\begin{equation*}
v\mathbf{\varpi }_{\tau }\mathbf{=}\left[
\begin{array}{cc}
0 & \overleftarrow{\varpi } \\
-\overleftarrow{\varpi }^{T} & v\mathbf{\Theta }%
\end{array}%
\right] \in \mathfrak{so}(m),~\overleftarrow{\varpi }\in \mathbb{R}^{m-1},~v%
\mathbf{\Theta \in }\mathfrak{so}(m-1).
\end{equation*}%
The components $h\mathbf{e}_{\parallel }$ and $h\overrightarrow{\mathbf{e}}%
_{\perp }$ correspond to the decomposition
\begin{equation*}
\mathbf{e}_{h\mathbf{Y}}=h\mathbf{g(\gamma }_{\tau },\mathbf{\gamma }_{%
\mathbf{l}}\mathbf{)e}_{h\mathbf{X}}+\mathbf{(\gamma }_{\tau })_{\perp
}\rfloor h\mathbf{e}_{\perp }
\end{equation*}%
into tangential and normal parts relative to $\mathbf{e}_{h\mathbf{X}}.$ In
a similar form, one considers $v\mathbf{e}_{\parallel }$ and $v%
\overleftarrow{\mathbf{e}}_{\perp }$ corresponding to the decomposition
\begin{equation*}
\mathbf{e}_{v\mathbf{Y}}=v\mathbf{g(\gamma }_{\tau },\mathbf{\gamma }_{%
\mathbf{l}}\mathbf{)e}_{v\mathbf{X}}+\mathbf{(\gamma }_{\tau })_{\perp
}\rfloor v\mathbf{e}_{\perp }.
\end{equation*}

Using the above stated matrix parametrizations, we get%
\begin{eqnarray}
\left[ \mathbf{e}_{h\mathbf{X}},\mathbf{e}_{h\mathbf{Y}}\right] &=&-\left[
\begin{array}{cc}
0 & 0 \\
0 & h\mathbf{e}_{\perp }%
\end{array}%
\right] \in \mathfrak{so}(n+1),  \label{aux41} \\
\mbox{ \ for \ }h\mathbf{e}_{\perp } &=&\left[
\begin{array}{cc}
0 & h\overrightarrow{\mathbf{e}}_{\perp } \\
-(h\overrightarrow{\mathbf{e}}_{\perp })^{T} & h\mathbf{0}%
\end{array}%
\right] \in \mathfrak{so}(n);  \notag \\
\left[ \mathbf{\Gamma }_{h\mathbf{Y}},\mathbf{e}_{h\mathbf{Y}}\right] &=&-%
\left[
\begin{array}{cc}
0 & \left( 0,\overrightarrow{\varpi }\right) \\
-\left( 0,\overrightarrow{\varpi }\right) ^{T} & 0%
\end{array}%
\right] \in h\mathfrak{p}_{\perp };  \notag \\
\left[ \mathbf{\Gamma }_{h\mathbf{X}},\mathbf{e}_{h\mathbf{Y}}\right] &=&-%
\left[
\begin{array}{cc}
0 & \left( -\overrightarrow{v}\cdot h\overrightarrow{\mathbf{e}}_{\perp },h%
\mathbf{e}_{\parallel }\overrightarrow{v}\right) \\
-\left( -\overrightarrow{v}\cdot h\overrightarrow{\mathbf{e}}_{\perp },h%
\mathbf{e}_{\parallel }\overrightarrow{v}\right) ^{T} & h\mathbf{0}%
\end{array}%
\right] \in h\mathfrak{p};  \notag
\end{eqnarray}%
and
\begin{eqnarray}
\left[ \mathbf{e}_{v\mathbf{X}},\mathbf{e}_{v\mathbf{Y}}\right] &=&-\left[
\begin{array}{cc}
0 & 0 \\
0 & v\mathbf{e}_{\perp }%
\end{array}%
\right] \in \mathfrak{so}(m+1),  \label{aux41a} \\
\mbox{ \ for \ }v\mathbf{e}_{\perp } &=&\left[
\begin{array}{cc}
0 & v\overrightarrow{\mathbf{e}}_{\perp } \\
-(v\overrightarrow{\mathbf{e}}_{\perp })^{T} & v\mathbf{0}%
\end{array}%
\right] \in \mathfrak{so}(m);  \notag \\
\left[ \mathbf{\Gamma }_{v\mathbf{Y}},\mathbf{e}_{v\mathbf{Y}}\right] &=&-%
\left[
\begin{array}{cc}
0 & \left( 0,\overleftarrow{\varpi }\right) \\
-\left( 0,\overleftarrow{\varpi }\right) ^{T} & 0%
\end{array}%
\right] \in v\mathfrak{p}_{\perp };  \notag \\
\left[ \mathbf{\Gamma }_{v\mathbf{X}},\mathbf{e}_{v\mathbf{Y}}\right] &=&-%
\left[
\begin{array}{cc}
0 & \left( -\overleftarrow{v}\cdot v\overleftarrow{\mathbf{e}}_{\perp },v%
\mathbf{e}_{\parallel }\overleftarrow{v}\right) \\
-\left( -\overleftarrow{v}\cdot v\overleftarrow{\mathbf{e}}_{\perp },v%
\mathbf{e}_{\parallel }\overleftarrow{v}\right) ^{T} & v\mathbf{0}%
\end{array}%
\right] \in v\mathfrak{p}.  \notag
\end{eqnarray}

We can use formulas (\ref{aux41}) and (\ref{aux41a}) in order to write the
structure equations (\ref{mtors}) and (\ref{mcurv}) in terms of N--adapted
curve flow operators soldered to the geometry Klein N--anholonomic spaces
using the relations (\ref{aux33}). One obtains respectively the $\mathbf{G}$%
--invariant N--adapted torsion and curvature generated by the canonical
d--connection,
\begin{equation}
\mathbf{T}(\gamma _{\tau },\gamma _{\mathbf{l}})=\left( \mathbf{D}_{\mathbf{X%
}}\gamma _{\tau }-\mathbf{D}_{\mathbf{Y}}\gamma _{\mathbf{l}}\right) \rfloor
\mathbf{e=D}_{\mathbf{X}}\mathbf{e}_{\mathbf{Y}}-\mathbf{D}_{\mathbf{Y}}%
\mathbf{e}_{\mathbf{X}}+\left[ \mathbf{\Gamma }_{\mathbf{X}},\mathbf{e}_{%
\mathbf{Y}}\right] -\left[ \mathbf{\Gamma }_{\mathbf{Y}},\mathbf{e}_{\mathbf{%
X}}\right]  \label{torscf}
\end{equation}%
and
\begin{equation}
\mathbf{R}(\gamma _{\tau },\gamma _{\mathbf{l}})\mathbf{e=}\left[ \mathbf{D}%
_{\mathbf{X}},\mathbf{D}_{\mathbf{Y}}\right] \mathbf{e=D}_{\mathbf{X}}%
\mathbf{\Gamma }_{\mathbf{Y}}-\mathbf{D}_{\mathbf{Y}}\mathbf{\Gamma }_{%
\mathbf{X}}+\left[ \mathbf{\Gamma }_{\mathbf{X}},\mathbf{\Gamma }_{\mathbf{Y}%
}\right]  \label{curvcf}
\end{equation}%
where $\mathbf{e}_{\mathbf{X}}\doteqdot \gamma _{\mathbf{l}}\rfloor \mathbf{%
e,}$ $\mathbf{e}_{\mathbf{Y}}\doteqdot \gamma _{\mathbf{\tau }}\rfloor
\mathbf{e,}$ $\mathbf{\Gamma }_{\mathbf{X}}\doteqdot \gamma _{\mathbf{l}%
}\rfloor \mathbf{\Gamma }$ and $\mathbf{\Gamma }_{\mathbf{Y}}\doteqdot
\gamma _{\mathbf{\tau }}\rfloor \mathbf{\Gamma .}$ The formulas (\ref{torscf}%
) and (\ref{curvcf}) are equivalent, respectively, to (\ref{dtors}) and (\ref%
{dcurv}). In general, $\mathbf{T}(\gamma _{\tau },\gamma _{\mathbf{l}})\neq
0 $ and $\mathbf{R}(\gamma _{\tau },\gamma _{\mathbf{l}})\mathbf{e}$ can not
be defined to have constant matrix coefficients with respect to a N--adapted
basis. For N--anholonomic spaces with dimensions $n=m,$ we have $^{\mathcal{E%
}}\mathbf{T}(\gamma _{\tau },\gamma _{\mathbf{l}})=0$ and $^{\mathcal{E}}%
\mathbf{R}(\gamma _{\tau },\gamma _{\mathbf{l}})\mathbf{e}$ defined by
constant, or vanishing, d--curvature coefficients (see discussions related
to formulas (\ref{dcurvtb}) and (\ref{candcontm})). For such cases, we can
consider the h-- and v--components of (\ref{torscf}) and (\ref{curvcf}) in a
similar manner as for symmetric Riemannian spaces but with the canonical
d--connection instead of the Levi Civita one. One obtains, respectively,%
\begin{eqnarray}
0 &=&\left( \mathbf{D}_{h\mathbf{X}}\gamma _{\tau }-\mathbf{D}_{h\mathbf{Y}%
}\gamma _{\mathbf{l}}\right) \rfloor h\mathbf{e}  \label{torseq} \\
&\mathbf{=}&\mathbf{D}_{h\mathbf{X}}\mathbf{e}_{h\mathbf{Y}}-\mathbf{D}_{h%
\mathbf{Y}}\mathbf{e}_{h\mathbf{X}}+\left[ \mathbf{L}_{h\mathbf{X}},\mathbf{e%
}_{h\mathbf{Y}}\right] -\left[ \mathbf{L}_{h\mathbf{Y}},\mathbf{e}_{h\mathbf{%
X}}\right] ;  \notag \\
0 &=&\left( \mathbf{D}_{v\mathbf{X}}\gamma _{\tau }-\mathbf{D}_{v\mathbf{Y}%
}\gamma _{\mathbf{l}}\right) \rfloor v\mathbf{e}  \notag \\
&\mathbf{=}&\mathbf{D}_{v\mathbf{X}}\mathbf{e}_{v\mathbf{Y}}-\mathbf{D}_{v%
\mathbf{Y}}\mathbf{e}_{v\mathbf{X}}+\left[ \mathbf{C}_{v\mathbf{X}},\mathbf{e%
}_{v\mathbf{Y}}\right] -\left[ \mathbf{C}_{v\mathbf{Y}},\mathbf{e}_{v\mathbf{%
X}}\right] ,  \notag
\end{eqnarray}%
and%
\begin{eqnarray}
h\mathbf{R}(\gamma _{\tau },\gamma _{\mathbf{l}})h\mathbf{e} &\mathbf{=}&%
\left[ \mathbf{D}_{h\mathbf{X}},\mathbf{D}_{h\mathbf{Y}}\right] h\mathbf{e=D}%
_{h\mathbf{X}}\mathbf{L}_{h\mathbf{Y}}-\mathbf{D}_{h\mathbf{Y}}\mathbf{L}_{h%
\mathbf{X}}+\left[ \mathbf{L}_{h\mathbf{X}},\mathbf{L}_{h\mathbf{Y}}\right]
\label{curveq} \\
v\mathbf{R}(\gamma _{\tau },\gamma _{\mathbf{l}})v\mathbf{e} &\mathbf{=}&%
\left[ \mathbf{D}_{v\mathbf{X}},\mathbf{D}_{v\mathbf{Y}}\right] v\mathbf{e=D}%
_{v\mathbf{X}}\mathbf{C}_{v\mathbf{Y}}-\mathbf{D}_{v\mathbf{Y}}\mathbf{C}_{v%
\mathbf{X}}+\left[ \mathbf{C}_{v\mathbf{X}},\mathbf{C}_{v\mathbf{Y}}\right] .
\notag
\end{eqnarray}

Following N--adapted curve flow parametrizations (\ref{aux41}) and (\ref%
{aux41a}), the equations (\ref{torseq}) and (\ref{curveq}) are written
\begin{eqnarray}
0 &=&\mathbf{D}_{h\mathbf{X}}h\mathbf{e}_{\parallel }+\overrightarrow{v}%
\cdot h\overrightarrow{\mathbf{e}}_{\perp },~0=\mathbf{D}_{v\mathbf{X}}v%
\mathbf{e}_{\parallel }+\overleftarrow{v}\cdot v\overleftarrow{\mathbf{e}}%
_{\perp },;  \label{torseqd} \\
~0 &=&\overrightarrow{\varpi }-h\mathbf{e}_{\parallel }\overrightarrow{v}+%
\mathbf{D}_{h\mathbf{X}}h\overrightarrow{\mathbf{e}}_{\perp },~0=%
\overleftarrow{\varpi }-v\mathbf{e}_{\parallel }\overleftarrow{v}+\mathbf{D}%
_{v\mathbf{X}}v\overleftarrow{\mathbf{e}}_{\perp };  \notag
\end{eqnarray}%
and%
\begin{eqnarray}
\mathbf{D}_{h\mathbf{X}}\overrightarrow{\varpi }-\mathbf{D}_{h\mathbf{Y}}%
\overrightarrow{v}+\overrightarrow{v}\rfloor h\mathbf{\Theta } &\mathbf{=}&h%
\overrightarrow{\mathbf{e}}_{\perp },~\mathbf{D}_{v\mathbf{X}}\overleftarrow{%
\varpi }-\mathbf{D}_{v\mathbf{Y}}\overleftarrow{v}+\overleftarrow{v}\rfloor v%
\mathbf{\Theta =}v\overleftarrow{\mathbf{e}}_{\perp };  \notag \\
\mathbf{D}_{h\mathbf{X}}h\mathbf{\Theta -}\overrightarrow{v}\otimes
\overrightarrow{\varpi }+\overrightarrow{\varpi }\otimes \overrightarrow{v}
&=&0,~\mathbf{D}_{v\mathbf{X}}v\mathbf{\Theta -}\overleftarrow{v}\otimes
\overleftarrow{\varpi }+\overleftarrow{\varpi }\otimes \overleftarrow{v}=0.
\label{curveqd}
\end{eqnarray}%
The tensor and interior products, for instance, for the h--components, are
defined in the form: $\otimes $ denotes the outer product of pairs of
vectors ($1\times n$ row matrices), producing $n\times n$ matrices $%
\overrightarrow{A}\otimes \overrightarrow{B}=\overrightarrow{A}^{T}%
\overrightarrow{B},$ and $\rfloor $ denotes multiplication of $n\times n$
matrices on vectors ($1\times n$ row matrices); one holds the properties $%
\overrightarrow{A}\rfloor \left( \overrightarrow{B}\otimes \overrightarrow{C}%
\right) =\left( \overrightarrow{A}\cdot \overrightarrow{B}\right)
\overrightarrow{C}$ which is the transpose of the standard matrix product on
column vectors, and $\left( \overrightarrow{B}\otimes \overrightarrow{C}%
\right) \overrightarrow{A}=\left( \overrightarrow{C}\cdot \overrightarrow{A}%
\right) \overrightarrow{B}.$ Here we note that similar formulas hold for the
v--components but, for instance, we have to change, correspondingly, $%
n\rightarrow m$ and $\overrightarrow{A}\rightarrow \overleftarrow{A}.$

The variables $\mathbf{e}_{\parallel }$ and $\mathbf{\Theta ,}$ written in
h-- and v--components, can be expressed corresponding in terms of variables $%
\overrightarrow{v},\overrightarrow{\varpi },h\overrightarrow{\mathbf{e}}%
_{\perp }$ and $\overleftarrow{v},\overleftarrow{\varpi },v\overleftarrow{%
\mathbf{e}}_{\perp }$ (see respectively the first two equations in (\ref%
{torseqd}) and the last two equations in (\ref{curveqd})),%
\begin{equation*}
h\mathbf{e}_{\parallel }=-\mathbf{D}_{h\mathbf{X}}^{-1}(\overrightarrow{v}%
\cdot h\overrightarrow{\mathbf{e}}_{\perp }),~v\mathbf{e}_{\parallel }=-%
\mathbf{D}_{v\mathbf{X}}^{-1}(\overleftarrow{v}\cdot v\overleftarrow{\mathbf{%
e}}_{\perp }),
\end{equation*}%
and%
\begin{equation*}
h\mathbf{\Theta =D}_{h\mathbf{X}}^{-1}\left( \overrightarrow{v}\otimes
\overrightarrow{\varpi }-\overrightarrow{\varpi }\otimes \overrightarrow{v}%
\right) ,~v\mathbf{\Theta =D}_{v\mathbf{X}}^{-1}\left( \overleftarrow{v}%
\otimes \overleftarrow{\varpi }-\overleftarrow{\varpi }\otimes
\overleftarrow{v}\right) .
\end{equation*}%
Substituting these values, correspondingly, in the last two equations in (%
\ref{torseqd}) and in the first two equations in (\ref{curveqd}), we express
\begin{equation*}
\overrightarrow{\varpi }=-\mathbf{D}_{h\mathbf{X}}h\overrightarrow{\mathbf{e}%
}_{\perp }-\mathbf{D}_{h\mathbf{X}}^{-1}(\overrightarrow{v}\cdot h%
\overrightarrow{\mathbf{e}}_{\perp })\overrightarrow{v},~\overleftarrow{%
\varpi }=-\mathbf{D}_{v\mathbf{X}}v\overleftarrow{\mathbf{e}}_{\perp }-%
\mathbf{D}_{v\mathbf{X}}^{-1}(\overleftarrow{v}\cdot v\overleftarrow{\mathbf{%
e}}_{\perp })\overleftarrow{v},
\end{equation*}%
contained in the h-- and v--flow equations respectively on $\overrightarrow{v%
}$ and $\overleftarrow{v},$ considered as scalar components when $\mathbf{D}%
_{h\mathbf{Y}}\overrightarrow{v}=\overrightarrow{v}_{\tau }$ and $\mathbf{D}%
_{h\mathbf{Y}}\overleftarrow{v}=\overleftarrow{v}_{\tau },$
\begin{eqnarray}
\overrightarrow{v}_{\tau } &=&\mathbf{D}_{h\mathbf{X}}\overrightarrow{\varpi
}-\overrightarrow{v}\rfloor \mathbf{D}_{h\mathbf{X}}^{-1}\left(
\overrightarrow{v}\otimes \overrightarrow{\varpi }-\overrightarrow{\varpi }%
\otimes \overrightarrow{v}\right) -\overrightarrow{R}h\overrightarrow{%
\mathbf{e}}_{\perp },  \label{floweq} \\
\overleftarrow{v}_{\tau } &=&\mathbf{D}_{v\mathbf{X}}\overleftarrow{\varpi }-%
\overleftarrow{v}\rfloor \mathbf{D}_{v\mathbf{X}}^{-1}\left( \overleftarrow{v%
}\otimes \overleftarrow{\varpi }-\overleftarrow{\varpi }\otimes
\overleftarrow{v}\right) -\overleftarrow{S}v\overleftarrow{\mathbf{e}}%
_{\perp },  \notag
\end{eqnarray}%
where the scalar curvatures of the canonical d--connection, $\overrightarrow{%
R}$ and $\overleftarrow{S}$ are defined by formulas (\ref{sdccurv}) in
Appendix. For symmetric Riemannian spaces like $SO(n+1)/SO(n)\simeq S^{n},$
the value $\overrightarrow{R}$ is just the scalar curvature $\chi =1,$ \ see %
\cite{anc2}. On N--anholonomic manifolds, it is possible that $%
\overrightarrow{R}$ and $\overleftarrow{S}$ are certain zero or nonzero
constants.

The above presented considerations consist the proof of

\begin{lemma}
On N--anholonomic spaces with constant curvature matrix coefficients for the
canonical d--connection, there are N--adapted Hamiltonian sympletic
operators,
\begin{equation}
h\mathcal{J}=\mathbf{D}_{h\mathbf{X}}+\mathbf{D}_{h\mathbf{X}}^{-1}\left(
\overrightarrow{v}\cdot \right) \overrightarrow{v}\mbox{ \ and \ }v\mathcal{J%
}=\mathbf{D}_{v\mathbf{X}}+\mathbf{D}_{v\mathbf{X}}^{-1}\left(
\overleftarrow{v}\cdot \right) \overleftarrow{v},  \label{sop}
\end{equation}%
and cosympletic operators%
\begin{equation}
h\mathcal{H}\doteqdot \mathbf{D}_{h\mathbf{X}}+\overrightarrow{v}\rfloor
\mathbf{D}_{h\mathbf{X}}^{-1}\left( \overrightarrow{v}\wedge \right)
\mbox{
\ and \ }v\mathcal{H}\doteqdot \mathbf{D}_{v\mathbf{X}}+\overleftarrow{v}%
\rfloor \mathbf{D}_{v\mathbf{X}}^{-1}\left( \overleftarrow{v}\wedge \right) ,
\label{csop}
\end{equation}%
where, for instance, $\overrightarrow{A}\wedge \overrightarrow{B}=%
\overrightarrow{A}\otimes \overrightarrow{B}-\overrightarrow{B}\otimes $ $%
\overrightarrow{A}.$\
\end{lemma}

The properties of operators (\ref{sop}) and (\ref{csop}) are defined by

\begin{theorem}
\label{mr1}The d--operators $\mathcal{J=}\left( h\mathcal{J},v\mathcal{J}%
\right) $ and $\mathcal{H=}\left( h\mathcal{H},v\mathcal{H}\right) $ $\ $are
respectively $\left( O(n-1),O(m-1)\right) $--invariant Hamiltonian sympletic
and cosympletic d--operators with respect to the Hamiltonian d--variables $%
\left( \overrightarrow{v},\overleftarrow{v}\right) .$ Such d--operators
defines the Hamiltonian form for the curve flow equations on N--anholonomic
manifolds with constant d--connection curvature: the h--flows are given by%
\begin{eqnarray}
\overrightarrow{v}_{\tau } &=&h\mathcal{H}\left( \overrightarrow{\varpi }%
\right) -\overrightarrow{R}~h\overrightarrow{\mathbf{e}}_{\perp }=h\mathfrak{%
R}\left( h\overrightarrow{\mathbf{e}}_{\perp }\right) -\overrightarrow{R}~h%
\overrightarrow{\mathbf{e}}_{\perp },  \notag \\
\overrightarrow{\varpi } &=&h\mathcal{J}\left( h\overrightarrow{\mathbf{e}}%
_{\perp }\right) ;  \label{hhfeq1}
\end{eqnarray}%
the v--flows are given by
\begin{eqnarray}
\overleftarrow{v}_{\tau } &=&v\mathcal{H}\left( \overleftarrow{\varpi }%
\right) -\overleftarrow{S}~v\overleftarrow{\mathbf{e}}_{\perp }=v\mathfrak{R}%
\left( v\overleftarrow{\mathbf{e}}_{\perp }\right) -\overleftarrow{S}~v%
\overleftarrow{\mathbf{e}}_{\perp },  \notag \\
\overleftarrow{\varpi } &=&v\mathcal{J}\left( v\overleftarrow{\mathbf{e}}%
_{\perp }\right) ,  \label{vhfeq1}
\end{eqnarray}%
where the so--called heriditary recursion d--operator has the respective h--
and v--components
\begin{equation}
h\mathfrak{R}=h\mathcal{H}\circ h\mathcal{J}\mbox{ \ and \ }v\mathfrak{R}=v%
\mathcal{H}\circ v\mathcal{J}.  \label{reqop}
\end{equation}
\end{theorem}

\begin{proof}
One follows from the Lemma and (\ref{floweq}). In a detailed form, for
holonomic structures, it is given in Ref. \cite{saw} and discussed in \cite%
{anc2}. The above considerations, in this section, consist a soldering of
certain classes of generalized Lagrange spaces with $\left(
O(n-1),O(m-1)\right)$--gauge symmetry to the geometry of Klein
N--anholonomic spaces.$\square $
\end{proof}

\subsection{Bi--Hamiltonian anholonomic curve flows and solitonic hierarchies%
}

Following a usual solitonic techniques, see details in Ref. \cite{anc1,anc2}%
, the recursion h--operator from (\ref{reqop}),%
\begin{eqnarray}
h\mathfrak{R} &=&\mathbf{D}_{h\mathbf{X}}\left( \mathbf{D}_{h\mathbf{X}}+%
\mathbf{D}_{h\mathbf{X}}^{-1}\left( \overrightarrow{v}\cdot \right)
\overrightarrow{v}\right) +\overrightarrow{v}\rfloor \mathbf{D}_{h\mathbf{X}%
}^{-1}\left( \overrightarrow{v}\wedge \mathbf{D}_{h\mathbf{X}}\right)
\label{reqoph} \\
&=&\mathbf{D}_{h\mathbf{X}}^{2}+|\mathbf{D}_{h\mathbf{X}}|^{2}+\mathbf{D}_{h%
\mathbf{X}}^{-1}\left( \overrightarrow{v}\cdot \right) \overrightarrow{v}_{%
\mathbf{l}}-\overrightarrow{v}\rfloor \mathbf{D}_{h\mathbf{X}}^{-1}(%
\overrightarrow{v}_{\mathbf{l}}\wedge ),  \notag
\end{eqnarray}%
generates a horizontal hierarchy of commuting Hamiltonian vector fields $h%
\overrightarrow{\mathbf{e}}_{\perp }^{(k)}$ starting from $h\overrightarrow{%
\mathbf{e}}_{\perp }^{(0)}=\overrightarrow{v}_{\mathbf{l}}$ given by the
infinitesimal generator of $\mathbf{l}$--translations in terms of arclength $%
\mathbf{l}$ along the curve (we use a boldface $\mathbf{l}$ in order to
emphasized that the curve is on a N--anholonomic manifold). A vertical
hierarchy of commuting vector fields $v\overleftarrow{\mathbf{e}}_{\perp
}^{(k)}$ starting from $v\overleftarrow{\mathbf{e}}_{\perp }^{(0)}$ $=%
\overleftarrow{v}_{\mathbf{l}}$ is generated by the recursion v--operator%
\begin{eqnarray}
v\mathfrak{R} &=&\mathbf{D}_{v\mathbf{X}}\left( \mathbf{D}_{v\mathbf{X}}+%
\mathbf{D}_{v\mathbf{X}}^{-1}\left( \overleftarrow{v}\cdot \right)
\overleftarrow{v}\right) +\overleftarrow{v}\rfloor \mathbf{D}_{v\mathbf{X}%
}^{-1}\left( \overleftarrow{v}\wedge \mathbf{D}_{v\mathbf{X}}\right)
\label{reqopv} \\
&=&\mathbf{D}_{v\mathbf{X}}^{2}+|\mathbf{D}_{v\mathbf{X}}|^{2}+\mathbf{D}_{v%
\mathbf{X}}^{-1}\left( \overleftarrow{v}\cdot \right) \overleftarrow{v}_{%
\mathbf{l}}-\overleftarrow{v}\rfloor \mathbf{D}_{v\mathbf{X}}^{-1}(%
\overleftarrow{v}_{\mathbf{l}}\wedge ).  \notag
\end{eqnarray}%
There are related hierarchies, generated by adjoint operators $\mathfrak{R}%
^{\ast }=(h\mathfrak{R}^{\ast },$ $v\mathfrak{R}^{\ast }),$ of involuntive
Hamiltonian h--covector fields $\overrightarrow{\varpi }^{(k)}=\delta \left(
hH^{(k)}\right) /\delta \overrightarrow{v}$ in terms of Hamiltonians $%
hH=hH^{(k)}(\overrightarrow{v},\overrightarrow{v}_{\mathbf{l}},%
\overrightarrow{v}_{2\mathbf{l}},...)$ starting from $\overrightarrow{\varpi
}^{(0)}=\overrightarrow{v},hH^{(0)}=\frac{1}{2}|\overrightarrow{v}|^{2}$ and
of involutive Hamiltonian v--covector fields $\overleftarrow{\varpi }%
^{(k)}=\delta \left( vH^{(k)}\right) /$ $\delta \overleftarrow{v}$ in terms
of Hamiltonians $vH=vH^{(k)}(\overleftarrow{v},\overleftarrow{v}_{\mathbf{l}%
},\overleftarrow{v}_{2\mathbf{l}},...)$ starting from $\overleftarrow{\varpi
}^{(0)}=\overleftarrow{v},vH^{(0)}=\frac{1}{2}|\overleftarrow{v}|^{2}.$ The
relations between hierarchies are established correspondingly by formulas%
\begin{equation*}
h\overrightarrow{\mathbf{e}}_{\perp }^{(k)}=h\mathcal{H}\left(
\overrightarrow{\varpi }^{(k)},\overrightarrow{\varpi }^{(k+1)}\right) =h%
\mathcal{J}\left( h\overrightarrow{\mathbf{e}}_{\perp }^{(k)}\right)
\end{equation*}%
and
\begin{equation*}
v\overleftarrow{\mathbf{e}}_{\perp }^{(k)}=v\mathcal{H}\left( \overleftarrow{%
\varpi }^{(k)},\overleftarrow{\varpi }^{(k+1)}\right) =v\mathcal{J}\left( v%
\overleftarrow{\mathbf{e}}_{\perp }^{(k)}\right) ,
\end{equation*}%
where $k=0,1,2,....$ All hierarchies (horizontal, vertical and their adjoint
ones) have a typical mKdV scaling symmetry, for instance, $\mathbf{%
l\rightarrow \lambda l}$ and $\overrightarrow{v}\rightarrow \mathbf{\lambda }%
^{-1}\overrightarrow{v}$ under which the values $h\overrightarrow{\mathbf{e}}%
_{\perp }^{(k)}$ and $hH^{(k)}$ have scaling weight $2+2k,$ while $%
\overrightarrow{\varpi }^{(k)}$ has scaling weight $1+2k.$

The above presented considerations prove

\begin{corollary}
\label{c2} There are N--adapted hierarchies of distinguished horizontal and
vertical commuting bi--Hamiltonian flows, correspondingly, on $%
\overrightarrow{v}$ and $\overleftarrow{v}$ associated to the recursion
d--operator (\ref{reqop}) given by $O(n-1)\oplus O(m-1)$ --invariant
d--vector evolution equations,%
\begin{eqnarray*}
\overrightarrow{v}_{\tau } &=&h\overrightarrow{\mathbf{e}}_{\perp }^{(k+1)}-%
\overrightarrow{R}~h\overrightarrow{\mathbf{e}}_{\perp }^{(k)}=h\mathcal{H}%
\left( \delta \left( hH^{(k,\overrightarrow{R})}\right) /\delta
\overrightarrow{v}\right) \\
&=&\left( h\mathcal{J}\right) ^{-1}\left( \delta \left( hH^{(k+1,%
\overrightarrow{R})}\right) /\delta \overrightarrow{v}\right)
\end{eqnarray*}%
with horizontal Hamiltonians $hH^{(k+1,\overrightarrow{R})}=hH^{(k+1,%
\overrightarrow{R})}-\overrightarrow{R}~hH^{(k,\overrightarrow{R})}$ and
\begin{eqnarray*}
\overleftarrow{v}_{\tau } &=&v\overleftarrow{\mathbf{e}}_{\perp }^{(k+1)}-%
\overleftarrow{S}~v\overleftarrow{\mathbf{e}}_{\perp }^{(k)}=v\mathcal{H}%
\left( \delta \left( vH^{(k,\overleftarrow{S})}\right) /\delta
\overleftarrow{v}\right) \\
&=&\left( v\mathcal{J}\right) ^{-1}\left( \delta \left( vH^{(k+1,%
\overleftarrow{S})}\right) /\delta \overleftarrow{v}\right)
\end{eqnarray*}%
with vertical Hamiltonians $vH^{(k+1,\overleftarrow{S})}=vH^{(k+1,%
\overleftarrow{S})}-\overleftarrow{S}~vH^{(k,\overleftarrow{S})},$ for $%
k=0,1,2,.....$ The d--operators $\mathcal{H}$ and $\mathcal{J}$ are
N--adapted and mutually compatible from which one can be constructed an
alternative (explicit) Hamilton d--operator $~^{a}\mathcal{H=H\circ J}$ $%
\circ \mathcal{H=}\mathfrak{R\circ }\mathcal{H}.$
\end{corollary}

\subsubsection{Formulation of the main theorem}

The main goal of this paper is to prove that for any regular Lagrange system
one can be defined naturally a N--adapted bi--Hamiltonian flow hierarchy
inducing anholonomic solitonic configurations.

\begin{theorem}
\label{mt} For any vector bundle with prescribed d--metric
structure, one can be defined a hierarchy of bi-Hamiltonian N--adapted flows
of curves $\gamma (\tau ,\mathbf{l})=h\gamma (\tau ,\mathbf{l})+v\gamma
(\tau ,\mathbf{l})$ described by geometric nonholonomic map equations. The $%
0 $ flows are defined as convective (travelling wave) maps%
\begin{equation}
\gamma _{\tau }=\gamma _{\mathbf{l}},\mbox{\ distinguished \ }\left( h\gamma
\right) _{\tau }=\left( h\gamma \right) _{h\mathbf{X}}\mbox{\ and \ }\left(
v\gamma \right) _{\tau }=\left( v\gamma \right) _{v\mathbf{X}}.
\label{trmap}
\end{equation}%
There are +1 flows defined as non--stretching mKdV maps%
\begin{eqnarray}
-\left( h\gamma \right) _{\tau } &=&\mathbf{D}_{h\mathbf{X}}^{2}\left(
h\gamma \right) _{h\mathbf{X}}+\frac{3}{2}\left| \mathbf{D}_{h\mathbf{X}%
}\left( h\gamma \right) _{h\mathbf{X}}\right| _{h\mathbf{g}}^{2}~\left(
h\gamma \right) _{h\mathbf{X}},  \label{1map} \\
-\left( v\gamma \right) _{\tau } &=&\mathbf{D}_{v\mathbf{X}}^{2}\left(
v\gamma \right) _{v\mathbf{X}}+\frac{3}{2}\left| \mathbf{D}_{v\mathbf{X}%
}\left( v\gamma \right) _{v\mathbf{X}}\right| _{v\mathbf{g}}^{2}~\left(
v\gamma \right) _{v\mathbf{X}},  \notag
\end{eqnarray}%
and the +2,...flows as higher order analogs. Finally, the -1 flows are
defined by the kernels of recursion operators (\ref{reqoph}) and (\ref%
{reqopv}) inducing non--stretching maps%
\begin{equation}
\mathbf{D}_{h\mathbf{Y}}\left( h\gamma \right) _{h\mathbf{X}}=0%
\mbox{\ and \
}\mathbf{D}_{v\mathbf{Y}}\left( v\gamma \right) _{v\mathbf{X}}=0.
\label{-1map}
\end{equation}
\end{theorem}

\begin{proof}
It is given in the next section \ref{ssp}.
\end{proof}

For similar constructions in gravity models with nontrivial torsion and
nonholonomic structure and related geometry of noncommutative/ super--
spaces and anholonomic spinors, it is important \cite{vncg,vsgg}

\begin{remark}
\label{rema}N--adapted hierarchies of bi--Hamiltonian operators and related
solitonic equations can be defined for $SU(n)\oplus SU(m)$ / $SO(n)\oplus
SO(m)$ symmetries like it was constructed in Ref. \cite{anc1} for the
Riemannian symmetric spaces. In this paper, we restrict our considerations
only for real nonholonomic models. Similar results, to those from the Theorem \ref{mt},
can be reformulated for unitary groups which may be very important in modern
quantum / (non)commutative gravity.
\end{remark}

Finally, it should be emphasized that a number of exact solutions in gravity
can be nonholonomically deformed in order to generate nonholonomic
hierachies of gravitational solitons of type (\ref{trmap}), (\ref{1map}) or (%
\ref{-1map}), which will be consider in our further publications.

\subsubsection{Proof of the main theorem}

\label{ssp}We provide a proof of Theorem \ref{mt} for the horizontal flows.
The approach is based on the method provided in Section 3 of Ref. \cite{anc1}
but in this work the Levi Civita connection on symmetric Riemannian spaces
is substituted by the horizontal components of the canonical d--connection
in a generalized Lagrange space with constant d--curvature coefficients. The
vertical constructions are similar but with respective changing of h--
variables / objects into v- variables/ objects.

One obtains a vector mKdV equation up to a convective term (can be absorbed
by redefinition of coordinates) defining the +1 flow for $h\overrightarrow{%
\mathbf{e}}_{\perp }=\overrightarrow{v}_{\mathbf{l}},$%
\begin{equation*}
\overrightarrow{v}_{\tau }=\overrightarrow{v}_{3\mathbf{l}}+\frac{3}{2}|%
\overrightarrow{v}|^{2}-\overrightarrow{R}~\overrightarrow{v}_{\mathbf{l}},
\end{equation*}%
when the $+(k+1)$ flow gives a vector mKdV equation of higher order $3+2k$
on $\overrightarrow{v}$ and there is a $0$ h--flow $\overrightarrow{v}_{\tau
}=\overrightarrow{v}_{\mathbf{l}}$ arising from $h\overrightarrow{\mathbf{e}}%
_{\perp }=0$ and $h\overrightarrow{\mathbf{e}}_{\parallel }=1$ belonging
outside the hierarchy generated by $h\mathfrak{R.}$ Such flows correspond to
N--adapted horizontal motions of the curve $\gamma (\tau ,\mathbf{l}%
)=h\gamma (\tau ,\mathbf{l})+v\gamma (\tau ,\mathbf{l}),$ given by
\begin{equation*}
\left( h\gamma \right) _{\tau }=f\left( \left( h\gamma \right) _{h\mathbf{X}%
},\mathbf{D}_{h\mathbf{X}}\left( h\gamma \right) _{h\mathbf{X}},\mathbf{D}_{h%
\mathbf{X}}^{2}\left( h\gamma \right) _{h\mathbf{X}},...\right)
\end{equation*}%
subject to the non--stretching condition $|\left( h\gamma \right) _{h\mathbf{%
X}}|_{h\mathbf{g}}=1,$ when the equation of motion is to be derived from the
identifications
\begin{equation*}
\left( h\gamma \right) _{\tau }\longleftrightarrow \mathbf{e}_{h\mathbf{Y}},%
\mathbf{D}_{h\mathbf{X}}\left( h\gamma \right) _{h\mathbf{X}%
}\longleftrightarrow \mathcal{D}_{h\mathbf{X}}\mathbf{e}_{h\mathbf{X}}=\left[
\mathbf{L}_{h\mathbf{X}},\mathbf{e}_{h\mathbf{X}}\right]
\end{equation*}%
and so on, which maps the constructions from the tangent space of the curve
to the space $h\mathfrak{p}.$ For such identifications, we have
\begin{eqnarray*}
\left[ \mathbf{L}_{h\mathbf{X}},\mathbf{e}_{h\mathbf{X}}\right] &=&-\left[
\begin{array}{cc}
0 & \left( 0,\overrightarrow{v}\right) \\
-\left( 0,\overrightarrow{v}\right) ^{T} & h\mathbf{0}%
\end{array}%
\right] \in h\mathfrak{p}, \\
\left[ \mathbf{L}_{h\mathbf{X}},\left[ \mathbf{L}_{h\mathbf{X}},\mathbf{e}_{h%
\mathbf{X}}\right] \right] &=&-\left[
\begin{array}{cc}
0 & \left( |\overrightarrow{v}|^{2},\overrightarrow{0}\right) \\
-\left( |\overrightarrow{v}|^{2},\overrightarrow{0}\right) ^{T} & h\mathbf{0}%
\end{array}%
\right]
\end{eqnarray*}%
and so on, see similar calculus in (\ref{aux41}). At the next step, stating
for the +1 h--flow
\begin{equation*}
h\overrightarrow{\mathbf{e}}_{\perp }=\overrightarrow{v}_{\mathbf{l}}%
\mbox{
and }h\overrightarrow{\mathbf{e}}_{\parallel }=-\mathbf{D}_{h\mathbf{X}%
}^{-1}\left( \overrightarrow{v}\cdot \overrightarrow{v}_{\mathbf{l}}\right)
=-\frac{1}{2}|\overrightarrow{v}|^{2},
\end{equation*}%
we compute
\begin{eqnarray*}
\mathbf{e}_{h\mathbf{Y}} &=&\left[
\begin{array}{cc}
0 & \left( h\mathbf{e}_{\parallel },h\overrightarrow{\mathbf{e}}_{\perp
}\right) \\
-\left( h\mathbf{e}_{\parallel },h\overrightarrow{\mathbf{e}}_{\perp
}\right) ^{T} & h\mathbf{0}%
\end{array}%
\right] \\
&=&-\frac{1}{2}|\overrightarrow{v}|^{2}\left[
\begin{array}{cc}
0 & \left( 1,\overrightarrow{\mathbf{0}}\right) \\
-\left( 0,\overrightarrow{\mathbf{0}}\right) ^{T} & h\mathbf{0}%
\end{array}%
\right] +\left[
\begin{array}{cc}
0 & \left( 0,\overrightarrow{v}_{h\mathbf{X}}\right) \\
-\left( 0,\overrightarrow{v}_{h\mathbf{X}}\right) ^{T} & h\mathbf{0}%
\end{array}%
\right] \\ &=& \mathbf{D}_{h\mathbf{X}}\left[ \mathbf{L}_{h\mathbf{X}},\mathbf{e}_{h%
\mathbf{X}}\right] +\frac{1}{2}\left[ \mathbf{L}_{h\mathbf{X}},\left[
\mathbf{L}_{h\mathbf{X}},\mathbf{e}_{h\mathbf{X}}\right] \right]
\\ &=& -\mathcal{D}_{h\mathbf{X}}\left[ \mathbf{L}_{h\mathbf{X}},\mathbf{e}_{h%
\mathbf{X}}\right] -\frac{3}{2}|\overrightarrow{v}|^{2}\mathbf{e}_{h\mathbf{X%
}}.
\end{eqnarray*}%
Following above presented identifications related to the first and second
terms, when
\begin{eqnarray*}
|\overrightarrow{v}|^{2} &=&<\left[ \mathbf{L}_{h\mathbf{X}},\mathbf{e}_{h%
\mathbf{X}}\right] ,\left[ \mathbf{L}_{h\mathbf{X}},\mathbf{e}_{h\mathbf{X}}%
\right] >_{h\mathfrak{p}}\longleftrightarrow h\mathbf{g}\left( \mathbf{D}_{h%
\mathbf{X}}\left( h\gamma \right) _{h\mathbf{X}},\mathbf{D}_{h\mathbf{X}%
}\left( h\gamma \right) _{h\mathbf{X}}\right) \\
&=&\left| \mathbf{D}_{h\mathbf{X}}\left( h\gamma \right) _{h\mathbf{X}%
}\right| _{h\mathbf{g}}^{2},
\end{eqnarray*}
we can identify $\mathcal{D}_{h\mathbf{X}}\left[ \mathbf{L}_{h\mathbf{X}},%
\mathbf{e}_{h\mathbf{X}}\right] $ to $\mathbf{D}_{h\mathbf{X}}^{2}\left(
h\gamma \right) _{h\mathbf{X}}$ and write
\begin{equation*}
-\mathbf{e}_{h\mathbf{Y}}\longleftrightarrow \mathbf{D}_{h\mathbf{X}%
}^{2}\left( h\gamma \right) _{h\mathbf{X}}+\frac{3}{2}\left| \mathbf{D}_{h%
\mathbf{X}}\left( h\gamma \right) _{h\mathbf{X}}\right| _{h\mathbf{g}%
}^{2}~\left( h\gamma \right) _{h\mathbf{X}}
\end{equation*}%
which is just the first equation (\ref{1map}) in the Theorem \ref{mt}
defining a non--stretching mKdV map h--equation induced by the h--part of
the canonical d--connection.

Using the adjoint representation $ad\left( \cdot \right) $ acting in the Lie
algebra $h\mathfrak{g}=h\mathfrak{p}\oplus \mathfrak{so}(n),$ with
\begin{equation*}
ad\left( \left[ \mathbf{L}_{h\mathbf{X}},\mathbf{e}_{h\mathbf{X}}\right]
\right) \mathbf{e}_{h\mathbf{X}}=\left[
\begin{array}{cc}
0 & \left( 0,\overrightarrow{\mathbf{0}}\right) \\
-\left( 0,\overrightarrow{\mathbf{0}}\right) ^{T} & \overrightarrow{\mathbf{v%
}}%
\end{array}%
\right] \in \mathfrak{so}(n+1),
\end{equation*}%
where%
\begin{equation*}
\overrightarrow{\mathbf{v}}=-\left[
\begin{array}{cc}
0 & \overrightarrow{v} \\
-\overrightarrow{v}^{T} & h\mathbf{0}%
\end{array}%
\in \mathfrak{so}(n)\right] ,
\end{equation*}%
and the derived (applying $ad\left( \left[ \mathbf{L}_{h\mathbf{X}},\mathbf{e%
}_{h\mathbf{X}}\right] \right) $ again )
\begin{equation*}
ad\left( \left[ \mathbf{L}_{h\mathbf{X}},\mathbf{e}_{h\mathbf{X}}\right]
\right) ^{2}\mathbf{e}_{h\mathbf{X}}=-|\overrightarrow{v}|^{2}\left[
\begin{array}{cc}
0 & \left( 1,\overrightarrow{\mathbf{0}}\right) \\
-\left( 1,\overrightarrow{\mathbf{0}}\right) ^{T} & \mathbf{0}%
\end{array}%
\right] =-|\overrightarrow{v}|^{2}\mathbf{e}_{h\mathbf{X}},
\end{equation*}%
the equation (\ref{1map}) can be represented in alternative form
\begin{equation*}
-\left( h\gamma \right) _{\tau }=\mathbf{D}_{h\mathbf{X}}^{2}\left( h\gamma
\right) _{h\mathbf{X}}-\frac{3}{2}\overrightarrow{R}^{-1}ad\left( \mathbf{D}%
_{h\mathbf{X}}\left( h\gamma \right) _{h\mathbf{X}}\right) ^{2}~\left(
h\gamma \right) _{h\mathbf{X}},
\end{equation*}%
which is more convenient for analysis of higher order flows on $%
\overrightarrow{v}$ subjected to higher--order geometric partial
differential equations. Here we note that the $0$ flow one $\overrightarrow{v%
}$ corresponds to just a convective (linear travelling h--wave but subjected
to certain nonholonomic constraints ) map equation (\ref{trmap}).

Now we consider a -1 flow contained in the h--hierarchy derived from the
property that $h\overrightarrow{\mathbf{e}}_{\perp }$ is annihilated by the
h--operator $h\mathcal{J}$ and mapped into $h\mathfrak{R}(h\overrightarrow{%
\mathbf{e}}_{\perp })=0.$This mean that $h\mathcal{J}(h\overrightarrow{%
\mathbf{e}}_{\perp })=\overrightarrow{\varpi }=0.$ Such properties together
with (\ref{auxaaa}) and equations (\ref{floweq}) imply $\mathbf{L}_{\tau }=0$
and hence $h\mathcal{D}_{\tau }\mathbf{e}_{h\mathbf{X}}=[\mathbf{L}_{\tau },%
\mathbf{e}_{h\mathbf{X}}]=0$ for $h\mathcal{D}_{\tau }=h\mathbf{D}_{\tau }+[%
\mathbf{L}_{\tau },\cdot ].$ We obtain the equation of motion for the
h--component of curve, $h\gamma (\tau ,\mathbf{l}),$ following the
correspondences $\mathbf{D}_{h\mathbf{Y}}\longleftrightarrow h\mathcal{D}%
_{\tau }$ and $h\gamma _{\mathbf{l}}\longleftrightarrow \mathbf{e}_{h\mathbf{%
X}},$%
\begin{equation*}
\mathbf{D}_{h\mathbf{Y}}\left( h\gamma (\tau ,\mathbf{l})\right) =0,
\end{equation*}%
which is just the first equation in (\ref{-1map}).

Finally, we note that the formulas for the v--components, stated by Theorem %
\ref{mt} can be derived in a similar form by respective substitution in the
the above proof of the h--operators and h--variables into v--ones, for
instance, $h\gamma \rightarrow v\gamma ,$ $h\overrightarrow{\mathbf{e}}%
_{\perp }\rightarrow v\overleftarrow{\mathbf{e}}_{\perp },$ $\overrightarrow{%
v}\rightarrow \overleftarrow{v},\overrightarrow{\varpi }\rightarrow
\overleftarrow{\varpi },\mathbf{D}_{h\mathbf{X}}\rightarrow \mathbf{D}_{v%
\mathbf{X}},$ $\mathbf{D}_{h\mathbf{Y}}\rightarrow \mathbf{D}_{v\mathbf{Y}},%
\mathbf{L\rightarrow C,}\overrightarrow{R}\rightarrow \overleftarrow{S},h%
\mathcal{D\rightarrow }v\mathcal{D},$ $h\mathfrak{R\rightarrow }v\mathfrak{R,%
}h\mathcal{J\rightarrow }v\mathcal{J}$,...

\subsection{Nonholonomic mKdV and SG hierarchies}

We consider explicit constructions when solitonic hierarchies are derived
following the conditions of Theorem \ref{mt}.

The h--flow and v--flow equations resulting from (\ref{-1map}) are%
\begin{equation}
\overrightarrow{v}_{\tau }=-\overrightarrow{R}h\overrightarrow{\mathbf{e}}%
_{\perp }\mbox{ \ and \ }\overleftarrow{v}_{\tau }=-\overleftarrow{S}v%
\overleftarrow{\mathbf{e}}_{\perp },  \label{deveq}
\end{equation}%
when, respectively,%
\begin{equation*}
0=\overrightarrow{\varpi }=-\mathbf{D}_{h\mathbf{X}}h\overrightarrow{\mathbf{%
e}}_{\perp }+h\mathbf{e}_{\parallel }\overrightarrow{v},~\mathbf{D}_{h%
\mathbf{X}}h\mathbf{e}_{\parallel }=h\overrightarrow{\mathbf{e}}_{\perp
}\cdot \overrightarrow{v}
\end{equation*}%
and
\begin{equation*}
0=\overleftarrow{\varpi }=-\mathbf{D}_{v\mathbf{X}}v\overleftarrow{\mathbf{e}%
}_{\perp }+v\mathbf{e}_{\parallel }\overleftarrow{v},~\mathbf{D}_{v\mathbf{X}%
}v\mathbf{e}_{\parallel }=v\overleftarrow{\mathbf{e}}_{\perp }\cdot
\overleftarrow{v}.
\end{equation*}%
The d--flow equations possess horizontal and vertical conservation laws%
\begin{equation*}
\mathbf{D}_{h\mathbf{X}}\left( (h\mathbf{e}_{\parallel })^{2}+|h%
\overrightarrow{\mathbf{e}}_{\perp }|^{2}\right) =0,
\end{equation*}%
for $(h\mathbf{e}_{\parallel })^{2}+|h\overrightarrow{\mathbf{e}}_{\perp
}|^{2}=<h\mathbf{e}_{\tau },h\mathbf{e}_{\tau }>_{h\mathfrak{p}}=|\left(
h\gamma \right) _{\tau }|_{h\mathbf{g}}^{2},$ and
\begin{equation*}
\mathbf{D}_{v\mathbf{Y}}\left( (v\mathbf{e}_{\parallel })^{2}+|v%
\overleftarrow{\mathbf{e}}_{\perp }|^{2}\right) =0,
\end{equation*}%
for $(v\mathbf{e}_{\parallel })^{2}+|v\overleftarrow{\mathbf{e}}_{\perp
}|^{2}=<v\mathbf{e}_{\tau },v\mathbf{e}_{\tau }>_{v\mathfrak{p}}=|\left(
v\gamma \right) _{\tau }|_{v\mathbf{g}}^{2}.$ This corresponds to
\begin{equation*}
\mathbf{D}_{h\mathbf{X}}|\left( h\gamma \right) _{\tau }|_{h\mathbf{g}}^{2}=0%
\mbox{ \ and \ }\mathbf{D}_{v\mathbf{X}}|\left( v\gamma \right) _{\tau }|_{v%
\mathbf{g}}^{2}=0.
\end{equation*}%
We note that the problem of formulating conservation laws on
N--anholo\-no\-mic spaces (in particular, on nonholonomic vector bundles) in
analyzed in Ref. \cite{vsgg}. In general, such laws are more sophisticate
than those on (semi) Riemannian spaces because ofn nonholonomic constraints
resulting in non--symmetric Ricci tensors and different types of identities.
But for the geometries modelled for dimensions $n=m$ with canonical
d--connections, we get similar h-- and v--components of the conservation law
equations as on symmetric Riemannian spaces.

It is possible to rescale conformally the variable $\tau $ in order to get $%
|\left( h\gamma \right) _{\tau }|_{h\mathbf{g}}^{2}$ $=1$ and (it could be
for other rescalling) $|\left( v\gamma \right) _{\tau }|_{v\mathbf{g}%
}^{2}=1, $ i.e. to have%
\begin{equation*}
(h\mathbf{e}_{\parallel })^{2}+|h\overrightarrow{\mathbf{e}}_{\perp }|^{2}=1%
\mbox{ \ and \ }(v\mathbf{e}_{\parallel })^{2}+|v\overleftarrow{\mathbf{e}}%
_{\perp }|^{2}=1.
\end{equation*}%
In this case, we can express $h\mathbf{e}_{\parallel }$ and $h%
\overrightarrow{\mathbf{e}}_{\perp }$ in terms of $\overrightarrow{v}$ and
its derivatives and, similarly, we can express $v\mathbf{e}_{\parallel }$
and $v\overleftarrow{\mathbf{e}}_{\perp }$ in terms of $\overleftarrow{v}$
and its derivatives, which follows from (\ref{deveq}). The N--adapted wave
map equations describing the -1 flows reduce to a system of two independent
nonlocal evolution equations for the h-- and v--components,%
\begin{equation*}
\overrightarrow{v}_{\tau }=-\mathbf{D}_{h\mathbf{X}}^{-1}\left( \sqrt{%
\overrightarrow{R}^{2}-|\overrightarrow{v}_{\tau }|^{2}}~\overrightarrow{v}%
\right) \mbox{ \ and \ }\overleftarrow{v}_{\tau }=-\mathbf{D}_{v\mathbf{X}%
}^{-1}\left( \sqrt{\overleftarrow{S}^{2}-|\overleftarrow{v}_{\tau }|^{2}}~%
\overleftarrow{v}\right) .
\end{equation*}%
For N--anholonomic spaces of constant scalar d--curvatures, we can rescale
the equations on $\tau $ to the case when the terms $\overrightarrow{R}^{2},%
\overleftarrow{S}^{2}=1,$ and the evolution equations transform into a
system of hyperbolic d--vector equations,%
\begin{equation}
\mathbf{D}_{h\mathbf{X}}(\overrightarrow{v}_{\tau })=-\sqrt{1-|%
\overrightarrow{v}_{\tau }|^{2}}~\overrightarrow{v}\mbox{ \ and \ }\mathbf{D}%
_{v\mathbf{X}}(\overleftarrow{v}_{\tau })=-\sqrt{1-|\overleftarrow{v}_{\tau
}|^{2}}~\overleftarrow{v},  \label{heq}
\end{equation}%
where $\mathbf{D}_{h\mathbf{X}}=\partial _{h\mathbf{l}}$ and $\mathbf{D}_{v%
\mathbf{X}}=\partial _{v\mathbf{l}}$ are usual partial derivatives on
direction $\mathbf{l=}h\mathbf{l+}v\mathbf{l}$ with $\overrightarrow{v}%
_{\tau }$ and $\overleftarrow{v}_{\tau }$ considered as scalar functions for
the covariant derivatives $\mathbf{D}_{h\mathbf{X}}$ and $\mathbf{D}_{v%
\mathbf{X}}$ defined by the canonical d--connection. It also follows that $h%
\overrightarrow{\mathbf{e}}_{\perp }$ and $v\overleftarrow{\mathbf{e}}%
_{\perp }$ obey corresponding vector sine--Gordon (SG) equations%
\begin{equation}
\left( \sqrt{(1-|h\overrightarrow{\mathbf{e}}_{\perp }|^{2})^{-1}}~\partial
_{h\mathbf{l}}(h\overrightarrow{\mathbf{e}}_{\perp })\right) _{\tau }=-h%
\overrightarrow{\mathbf{e}}_{\perp }  \label{sgeh}
\end{equation}%
and
\begin{equation}
\left( \sqrt{(1-|v\overleftarrow{\mathbf{e}}_{\perp }|^{2})^{-1}}~\partial
_{v\mathbf{l}}(v\overleftarrow{\mathbf{e}}_{\perp })\right) _{\tau }=-v%
\overleftarrow{\mathbf{e}}_{\perp }.  \label{sgev}
\end{equation}

The above presented formulas and Corollary \ref{c2} imply

\begin{conclusion}
The recursion d--operator $\mathfrak{R}=(h\mathfrak{R,}h\mathfrak{R})$ (\ref%
{reqop}), see (\ref{reqoph}) and (\ref{reqopv}), generates two hierarchies
of vector mKdV symmetries: the first one is horizontal,
\begin{eqnarray}
\overrightarrow{v}_{\tau }^{(0)} &=&\overrightarrow{v}_{h\mathbf{l}},~%
\overrightarrow{v}_{\tau }^{(1)}=h\mathfrak{R}(\overrightarrow{v}_{h\mathbf{l%
}})=\overrightarrow{v}_{3h\mathbf{l}}+\frac{3}{2}|\overrightarrow{v}|^{2}~%
\overrightarrow{v}_{h\mathbf{l}},  \label{mkdv1a} \\
\overrightarrow{v}_{\tau }^{(2)} &=&h\mathfrak{R}^{2}(\overrightarrow{v}_{h%
\mathbf{l}})=\overrightarrow{v}_{5h\mathbf{l}}+\frac{5}{2}\left( |%
\overrightarrow{v}|^{2}~\overrightarrow{v}_{2h\mathbf{l}}\right) _{h\mathbf{l%
}}  \notag \\
&&+\frac{5}{2}\left( (|\overrightarrow{v}|^{2})_{h\mathbf{l~}h\mathbf{l}}+|%
\overrightarrow{v}_{h\mathbf{l}}|^{2}+\frac{3}{4}|\overrightarrow{v}%
|^{4}\right) ~\overrightarrow{v}_{h\mathbf{l}}-\frac{1}{2}|\overrightarrow{v}%
_{h\mathbf{l}}|^{2}~\overrightarrow{v},  \notag \\
&&...,  \notag
\end{eqnarray}%
with all such terms commuting with the -1 flow
\begin{equation}
(\overrightarrow{v}_{\tau })^{-1}=h\overrightarrow{\mathbf{e}}_{\perp }
\label{mkdv1b}
\end{equation}%
associated to the vector SG equation (\ref{sgeh}); the second one is
vertical,
\begin{eqnarray}
\overleftarrow{v}_{\tau }^{(0)} &=&\overleftarrow{v}_{v\mathbf{l}},~%
\overleftarrow{v}_{\tau }^{(1)}=v\mathfrak{R}(\overleftarrow{v}_{v\mathbf{l}%
})=\overleftarrow{v}_{3v\mathbf{l}}+\frac{3}{2}|\overleftarrow{v}|^{2}~%
\overleftarrow{v}_{v\mathbf{l}},  \label{mkdv2a} \\
\overleftarrow{v}_{\tau }^{(2)} &=&v\mathfrak{R}^{2}(\overleftarrow{v}_{v%
\mathbf{l}})=\overleftarrow{v}_{5v\mathbf{l}}+\frac{5}{2}\left( |%
\overleftarrow{v}|^{2}~\overleftarrow{v}_{2v\mathbf{l}}\right) _{v\mathbf{l}}
\notag \\
&&+\frac{5}{2}\left( (|\overleftarrow{v}|^{2})_{v\mathbf{l~}v\mathbf{l}}+|%
\overleftarrow{v}_{v\mathbf{l}}|^{2}+\frac{3}{4}|\overleftarrow{v}%
|^{4}\right) ~\overleftarrow{v}_{v\mathbf{l}}-\frac{1}{2}|\overleftarrow{v}%
_{v\mathbf{l}}|^{2}~\overleftarrow{v},  \notag \\
&&...,  \notag
\end{eqnarray}%
with all such terms commuting with the -1 flow
\begin{equation}
(\overleftarrow{v}_{\tau })^{-1}=v\overleftarrow{\mathbf{e}}_{\perp }
\label{mkdv2b}
\end{equation}%
associated to the vector SG equation (\ref{sgev}).
\end{conclusion}

In its turn, using the above Conclusion, we derive that the adjoint
d--operator $\mathfrak{R}^{\ast }=\mathcal{J\circ H}$ generates a horizontal
hierarchy of Hamiltonians,%
\begin{eqnarray}
hH^{(0)} &=&\frac{1}{2}|\overrightarrow{v}|^{2},~hH^{(1)}=-\frac{1}{2}|%
\overrightarrow{v}_{h\mathbf{l}}|^{2}+\frac{1}{8}|\overrightarrow{v}|^{4},
\label{hhh} \\
hH^{(2)} &=&\frac{1}{2}|\overrightarrow{v}_{2h\mathbf{l}}|^{2}-\frac{3}{4}|%
\overrightarrow{v}|^{2}~|\overrightarrow{v}_{h\mathbf{l}}|^{2}-\frac{1}{2}%
\left( \overrightarrow{v}\cdot \overrightarrow{v}_{h\mathbf{l}}\right) +%
\frac{1}{16}|\overrightarrow{v}|^{6},...,  \notag
\end{eqnarray}%
and vertical hierarchy of Hamiltonians%
\begin{eqnarray}
vH^{(0)} &=&\frac{1}{2}|\overleftarrow{v}|^{2},~vH^{(1)}=-\frac{1}{2}|%
\overleftarrow{v}_{v\mathbf{l}}|^{2}+\frac{1}{8}|\overleftarrow{v}|^{4},
\label{hhv} \\
vH^{(2)} &=&\frac{1}{2}|\overleftarrow{v}_{2v\mathbf{l}}|^{2}-\frac{3}{4}|%
\overleftarrow{v}|^{2}~|\overleftarrow{v}_{v\mathbf{l}}|^{2}-\frac{1}{2}%
\left( \overleftarrow{v}\cdot \overleftarrow{v}_{v\mathbf{l}}\right) +\frac{1%
}{16}|\overleftarrow{v}|^{6},...,  \notag
\end{eqnarray}%
all of which are conserved densities for respective horizontal and vertical
-1 flows and determining higher conservation laws for the corresponding
hyperolic equations (\ref{sgeh}) and (\ref{sgev}).

The above presented horizontal equations (\ref{sgeh}), (\ref{mkdv1a}), (\ref%
{mkdv1b}) and (\ref{hhh}) and of vertical equations (\ref{sgev}), (\ref%
{mkdv2a}), (\ref{mkdv2b}) and (\ref{hhv}) have similar mKdV scaling
symmetries but on different parameters $\lambda _{h}$ and $\lambda _{v}$
because, in general, there are two independent values of scalar curvatures $%
\overrightarrow{R}$ and $\overleftarrow{S},$ see (\ref{sdccurv}). The
horizontal scaling symmetries are $h\mathbf{l\rightarrow }\lambda _{h}h%
\mathbf{l,}\overrightarrow{v}\rightarrow \left( \lambda _{h}\right) ^{-1}%
\overrightarrow{v}$ and $\tau \rightarrow \left( \lambda _{h}\right)
^{1+2k}, $ for $k=-1,0,1,2,...$ For the vertical scaling symmetries, one has
$v\mathbf{l\rightarrow }\lambda _{v}v\mathbf{l,}\overleftarrow{v}\rightarrow
\left( \lambda _{v}\right) ^{-1}\overleftarrow{v}$ and $\tau \rightarrow
\left( \lambda _{v}\right) ^{1+2k},$ for $k=-1,0,1,2,...$

Finally, we consider again the Remark \ref{rema} stating that similar
results (proved in Section 4) can be alternatively derived for unitary
groups with complex variables. It is really so, but the generated
bi--Hamiltonian and solitonic horizontal and vertical hierarchies with
unitary gauge symmetry are different from those defined for real orthogonal
groups; for holonomic spaces this is demonstrated in Section 4 of Ref. \cite%
{anc1}. This distinguishes substantially the models of gauge gravity with
structure groups like the unitary one from those with orthogonal groups.

\section{Conclusion}

In this paper, the geometry of (semi) Riemannian spaces was encoded in
nonholonomic hierarchies of bi--Hamiltonian structures and related solitonic
equations derived for curve flows on tangent spaces. The local algebraic
structure of modelled nonholonomic spaces is defined by the dimensions of
the base and typical fiber subspaces. If such
subspaces are Riemannian symmetric manifolds, respectively, of dimensions $n$
and $m,$ their geometric properties are exhausted by the geometry of
distinguished Lie groups $\mathbf{G}=GO(n)\oplus $ $GO(m)$ and $\mathbf{G}%
=SU(n)\oplus $ $SU(m)$ and the geometry of nonlinear connections on such
 vector bundles. This can be formulated equivalently in terms of
geometric objects on couples of Klein spaces. The bi--Hamiltonian and
related solitonic (of type mKdV and SG) hierarchies are generated naturally
by wave map equations and recursion operators associated to the horizontal
and vertical flows of curves on such spaces.

We proved that for any (semi) Riemanninan metric on a base manifold $M$ it
is possible to define canonical geometric object and their nonholonomic deformations
on  tangent bundles.   The curvature matrix, with respect to
the correspondingly adapted frames, can be constructed to posses  constant
coefficients. For such configurations, we can apply the former methods
elaborated for symmetric Riemannian spaces in order to generate curve flow
-- solitonic hierarchies.

Finally, we note that curve flow -- solitonic hierarchies can be constructed
in a similar manner for exact solutions of Einstein--Yang--Mills--Dirac
equations, derived following the anholonomic frame method, in noncommutative
generalizations of gravity and geometry and possible quantum models based on
nonholonomic Lagrange--Fedosov manifolds. We are continuing to work in such
directions.

\vskip5pt

\textbf{Acknowledgement: }The author is grateful to A. Bejancu for very
important references on the geometry of nonholonomic manifolds.

\appendix

\section{Some Local Formulas}

There are outlined some local results from geometry of nonlinear connections
(see Refs. \cite{ma1,ma2,vncg,vsgg} for proofs and details). There are two
types of preferred linear connections uniquely determined by a generic
off--diagonal metric structure with $n+m$ splitting, see $\mathbf{g}=g\oplus
_{N}h$ (\ref{m1}):

\begin{enumerate}
\item The Levi Civita connection $\nabla =\{\Gamma _{\beta \gamma }^{\alpha
}\}$ is by definition torsionless, $~\ _{\shortmid }\mathcal{T}=0,$ and
satisfies the metric compatibility condition$,\nabla \mathbf{g}=0.$

\item The canonical d--connection $\widehat{\mathbf{\Gamma }}_{\ \alpha
\beta }^{\gamma }=\left( \widehat{L}_{jk}^{i},\widehat{L}_{bk}^{a},\widehat{C%
}_{jc}^{i},\widehat{C}_{bc}^{a}\right) $ is also metric compatible, i. e. $%
\widehat{\mathbf{D}}\mathbf{g}=0,$ but the torsion vanishes only on h-- and
v--subspaces, i.e. $\widehat{T}_{jk}^{i}=0$ and $\widehat{T}_{bc}^{a}=0,$
for certain nontrivial values of $\widehat{T}_{ja}^{i},\widehat{T}_{bi}^{a},%
\widehat{T}_{ji}^{a}.$ For simplicity, we omit hats on symbols and write, for
simplicity, $L_{jk}^{i}$ instead of $\widehat{L}_{jk}^{i},$ $T_{ja}^{i}$
instead of $\widehat{T}_{ja}^{i}$ and so on...but preserve the general
symbols $\widehat{\mathbf{D}}$ and $\widehat{\mathbf{\Gamma }}_{\ \alpha
\beta }^{\gamma }.$
\end{enumerate}

By a straightforward calculus with respect to N--adapted frames (\ref{dder})
and (\ref{ddif}), one can verify that the requested properties for $\widehat{%
\mathbf{D}}$ on $\mathbf{E}$ are satisfied if
\begin{eqnarray}
L_{jk}^{i} &=&\frac{1}{2}g^{ir}\left( \mathbf{e}_{k}g_{jr}+\mathbf{e}%
_{j}g_{kr}-\mathbf{e}_{r}g_{jk}\right) ,  \label{candcon} \\
L_{bk}^{a} &=&e_{b}(N_{k}^{a})+\frac{1}{2}h^{ac}\left( \mathbf{e}%
_{k}h_{bc}-h_{dc}\ e_{b}N_{k}^{d}-h_{db}\ e_{c}N_{k}^{d}\right) ,  \notag \\
C_{jc}^{i} &=&\frac{1}{2}g^{ik}e_{c}g_{jk},\ C_{bc}^{a}=\frac{1}{2}%
h^{ad}\left( e_{c}h_{bd}+e_{c}h_{cd}-e_{d}h_{bc}\right) .  \notag
\end{eqnarray}%
For $\mathbf{E}=TM,$ the canonical d--connection $\ \mathbf{\tilde{D}}=(h%
\tilde{D},v\tilde{D})$ can be defined in torsionless form\footnote{%
i.e. it has the same coefficients as the Levi Civita connection with respect
to N--elongated bases (\ref{dder}) and (\ref{ddif})} with the coefficients $%
\Gamma _{\ \beta \gamma }^{\alpha }=(L_{\ jk}^{i},L_{bc}^{a}),$
\begin{eqnarray}
L_{\ jk}^{i} &=&\frac{1}{2}g^{ih}(\mathbf{e}_{k}g_{jh}+\mathbf{e}_{j}g_{kh}-%
\mathbf{e}_{h}g_{jk}),  \label{candcontm} \\
C_{\ bc}^{a} &=&\frac{1}{2}h^{ae}(e_{c}h_{be}+e_{b}h_{ce}-e_{e}h_{bc}).
\notag
\end{eqnarray}

The curvature of a d--connection $\mathbf{D,}$
\begin{equation}
\mathcal{R}_{~\beta }^{\alpha }\doteqdot \mathbf{D\Gamma }_{\ \beta
}^{\alpha }=d\mathbf{\Gamma }_{\ \beta }^{\alpha }-\mathbf{\Gamma }_{\ \beta
}^{\gamma }\wedge \mathbf{\Gamma }_{\ \gamma }^{\alpha },  \label{curv}
\end{equation}%
splits into six types of N--adapted components with respect to (\ref{dder})
and (\ref{ddif}),
\begin{equation*}
\mathbf{R}_{~\beta \gamma \delta }^{\alpha }=\left(
R_{~hjk}^{i},R_{~bjk}^{a},P_{~hja}^{i},P_{~bja}^{c},S_{~jbc}^{i},S_{~bdc}^{a}\right) ,
\end{equation*}%
\begin{eqnarray}
R_{\ hjk}^{i} &=&\mathbf{e}_{k}L_{\ hj}^{i}-\mathbf{e}_{j}L_{\ hk}^{i}+L_{\
hj}^{m}L_{\ mk}^{i}-L_{\ hk}^{m}L_{\ mj}^{i}-C_{\ ha}^{i}\Omega _{\ kj}^{a},
 \label{dcurv} \\
R_{\ bjk}^{a} &=&\mathbf{e}_{k}L_{\ bj}^{a}-\mathbf{e}_{j}L_{\ bk}^{a}+L_{\
bj}^{c}L_{\ ck}^{a}-L_{\ bk}^{c}L_{\ cj}^{a}-C_{\ bc}^{a}\Omega _{\ kj}^{c},
\notag \\
P_{\ jka}^{i} &=&e_{a}L_{\ jk}^{i}-D_{k}C_{\ ja}^{i}+C_{\ jb}^{i}T_{\
ka}^{b},~P_{\ bka}^{c}=e_{a}L_{\ bk}^{c}-D_{k}C_{\ ba}^{c}+C_{\ bd}^{c}T_{\
ka}^{c},   \notag \\
S_{\ jbc}^{i} &=&e_{c}C_{\ jb}^{i}-e_{b}C_{\ jc}^{i}+C_{\ jb}^{h}C_{\
hc}^{i}-C_{\ jc}^{h}C_{\ hb}^{i},  \notag \\
S_{\ bcd}^{a} &=&e_{d}C_{\ bc}^{a}-e_{c}C_{\ bd}^{a}+C_{\ bc}^{e}C_{\
ed}^{a}-C_{\ bd}^{e}C_{\ ec}^{a}.  \notag
\end{eqnarray}

Contracting respectively the components, $\mathbf{R}_{\alpha \beta
}\doteqdot \mathbf{R}_{\ \alpha \beta \tau }^{\tau },$ one computes the h-
v--components of the Ricci d--tensor (there are four N--adapted components)
\begin{equation}
R_{ij}\doteqdot R_{\ ijk}^{k},\ \ R_{ia}\doteqdot -P_{\ ika}^{k},\
R_{ai}\doteqdot P_{\ aib}^{b},\ S_{ab}\doteqdot S_{\ abc}^{c}.
\label{dricci}
\end{equation}%
The scalar curvature is defined by contracting the Ricci d--tensor with the
inverse metric $\mathbf{g}^{\alpha \beta },$
\begin{equation}
\overleftrightarrow{\mathbf{R}}\doteqdot \mathbf{g}^{\alpha \beta }\mathbf{R}%
_{\alpha \beta }=g^{ij}R_{ij}+h^{ab}S_{ab}=\overrightarrow{R}+\overleftarrow{%
S}.  \label{sdccurv}
\end{equation}

If $\mathbf{E=}TM,$ there are only three classes of d--curvatures,%
\begin{eqnarray}
R_{\ hjk}^{i} &=&\mathbf{e}_{k}L_{\ hj}^{i}-\mathbf{e}_{j}L_{\ hk}^{i}+L_{\
hj}^{m}L_{\ mk}^{i}-L_{\ hk}^{m}L_{\ mj}^{i}-C_{\ ha}^{i}\Omega _{\ kj}^{a},
\label{dcurvtb} \\
P_{\ jka}^{i} &=&e_{a}L_{\ jk}^{i}-\mathbf{D}_{k}C_{\ ja}^{i}+C_{\
jb}^{i}T_{\ ka}^{b}, \notag \\
S_{\ bcd}^{a}&=&e_{d}C_{\ bc}^{a}-e_{c}C_{\ bd}^{a}+C_{\
bc}^{e}C_{\ ed}^{a}-C_{\ bd}^{e}C_{\ ec}^{a},  \notag
\end{eqnarray}%
where all indices $a,b,...,i,j,...$ run the same values and, for
instance, $C_{\ bc}^{e}\to $ $C_{\ jk}^{i},...$

\end{document}